\documentclass[preprint]{elsarticle}
                    
\usepackage{graphicx}
\usepackage{graphics}
\usepackage{epsfig}
\epsfclipon

\usepackage{epsfig,bm}
\usepackage{latexsym}
\usepackage{amssymb,amsmath}

\usepackage{epsfig,bm}
\usepackage{latexsym}
\usepackage{amssymb,amsmath}

\usepackage[small]{caption}

\usepackage{graphicx} 
\usepackage{amsmath}
\usepackage{amsfonts}
\usepackage{amssymb}

\newcommand{\be}{\begin{equation}}
\newcommand{\ee}{\end{equation}}
\newcommand{\bea}{\begin{eqnarray}}
\newcommand{\eea}{\end{eqnarray}}

\def\lsim{\mathrel{\rlap{\lower4pt\hbox{\hskip1pt$\sim$}}
    \raise1pt\hbox{$<$}}}                
\def\gsim{\mathrel{\rlap{\lower4pt\hbox{\hskip1pt$\sim$}}
    \raise1pt\hbox{$>$}}}                

\def\revise#1       {\raisebox{-0em}{\rule{3pt}{1em}}%
                     \marginpar{\raisebox{.5em}{\vrule width3pt\
                     \vrule width0pt height 0pt depth0.5em
                     \hbox to 0cm{\hspace{0cm}{%
                     \parbox[t]{4em}{\raggedright\footnotesize{#1}}}\hss}}}}

\def\sqr#1#2{{\vcenter{\vbox{\hrule height.#2pt
 \hbox{\vrule width.#2pt height#1pt \kern#1pt
 \vrule width.#2pt}\hrule height.#2pt}}}}

\def\aa1{\phi}
\def\cc1{\psi}

\newcommand{\eq}{\begin{equation}}
\newcommand{\eqx}{\end{equation}}
\newcommand{\eqn}{\begin{eqnarray}}
\newcommand{\eqnx}{\end{eqnarray}}

\def\Vint{V_{\mathrm{int}}}

\catcode`\@=12

\journal{Nuclear Physics B}

\begin{document}

\begin{frontmatter}

\title{A New Formulation of the Initial Value Problem for Nonlocal Theories}

\author{Neil Barnaby}
\ead{barnaby@cita.utoronto.ca}
\address{Canadian Institute for Theoretical Astrophysics, University of Toronto,\\
60 St. George St., Toronto, ON M5S 3H8, Canada}

\begin{abstract}
There are a number of reasons to entertain the possibility that locality is violated on microscopic scales, for example
through the presence of an infinite series of higher derivatives in the fundamental equations of motion.  This type of nonlocality
leads to improved UV behaviour, novel cosmological dynamics and is a generic prediction of string theory.  On the other hand,
fundamentally nonlocal models are fraught with complications, including instabilities and complications in setting up the initial value
problem.  We study the structure of the initial value problem in an interesting class of nonlocal models.  We advocate
a novel new formulation wherein the Cauchy surface is ``smeared out'' over the underlying scale of nonlocality, so that the the usual notion
of initial data at $t=0$ is replaced with an ``initial function'' defined over $-M^{-1} \leq t \leq 0$ where $M$ is the underlying scale of nonlocality.  
Focusing on some specific examples from string theory and cosmology, we show that this mathematical re-formulation has surprising implications
for the well-known stability problem.  For D-brane decay in a linear dilaton background, we are able to show that the unstable directions in phase space
cannot be accessed starting from a physically sensible initial function.  Previous examples of unstable solutions in this model therefore correspond
to unphysical initial conditions, an observation which is obfuscated in the old formulation of the initial value problem.   We also discuss implication of 
this approach for nonlocal cosmological models.
\end{abstract}



\end{frontmatter}

\section{Introduction}

Nearly all dynamical systems encountered in physics are local.  Time evolution is usually described by differential equations 
involving at most two time derivatives, and the $t>0$ dynamics depend only on a small amount of data specifying the 
state of the system at $t=0$.  There are, however, a number of reasons to entertain the possibility that locality is violated on microscopic scales,
for example through the appearance of an infinite series of higher derivatives in the fundamental equations of motion.  
This type of nonlocality is a generic feature of many particle physics models including string theory \cite{sft_rev}, field theories living on noncommutative 
space-time \cite{nc_rev} and also models with a minimal length scale \cite{l_min}.  In fact, it has been argued that \emph{any} Quantum Field Theory 
(QFT) of gravity must be nonlocal at the most fundamental level \cite{woodard}.

The underlying nonlocality of string theory manifests itself strikingly through the infinite derivative structure of String Field Theory \cite{sft} (SFT) and also 
toy models such as the $p$-adic string \cite{padic,zwiebach} or strings quantized on a random lattice \cite{lattice}.  These constructions make it possible 
to study both nonperturbative and off-shell string physics, and have therefore played a key role in describing the decay of unstable D-brane 
configurations (tachyon condensation).  The dynamics of tachyon condensation \cite{sen_rev} are relevant for both formal string theory and also 
cosmology, because brane/anti-brane annihilation \cite{annihilation} marks the endpoint of popular D-brane inflation models \cite{brane_inf}.

Aside from their ``bottom up'' motivations, field theories with infinitely many derivatives are interesting due to improved Ultra-Violet (UV) behaviour 
\cite{pais,finite} and novel cosmological dynamics \cite{woodard_cosmo,koivisto,phantom,lidsey,mulryne,tirtho,justin,padic_inf,ng1,ng2}.  In 
\cite{padic_inf,ng1} it was shown that nonlocal theories can support inflation even on a very steep potential, alleviating the fine 
tuning problems that plague inflationary model building.  This mechanism leads to a robust prediction for large nongaussianity in the Cosmic Microwave 
Background (CMB) which will be falsifiable (or verifiable!) with Planck \cite{ng1,ng2}.

In spite of these motivations, physical applications of nonlocal field theories are fraught with complications.  Perhaps the most serious concerns are the
following. 
\begin{enumerate}
  \item {\bf Stability:} If the equation of motion admits more than two initial data then the ``extra'' degrees of freedom can be interpreted
  as physical excitations which carry wrong-sign kinetic energy and render the Hamiltonian unbounded from below \cite{woodard,pais,niky1,nl_rev}.  The 
  quantum theory contains ghosts, while the classical theory is plagued by Ostrogradski (higher derivative) instabilities.  
  \item {\bf Predictivity:} If the equation of motion requires \emph{infinitely} many initial data, then there is an independent concern.  By suitable choice
  of the infinite free parameters of the solution it might be possible to arrange for nearly \emph{any} time dependence over an arbitrarily long interval.
  In this case the initial value problem would be completely bereft of predictivity.  (Throughout this paper we understand the word ``predictivity''
  in this sense.)
\end{enumerate}
Clearly, very much is at stake.  There is evidence that \emph{both} of these problems might manifest themselves in string theory \cite{woodard}, 
for example through the wild oscillatory dynamics of tachyon condensation in flat space \cite{zwiebach} (however, see \cite{no_prob}). 

As emphasized in \cite{niky1}, the presence of ghost-like instabilities in a nonlocal QFT is intimately related to the structure of the initial value problem
for the classical equations of motion.  In this work, we consider a new formulation of the initial value problem for a broad class of nonlocal theories which
incorporate dissipation \cite{simeon,BMNR,delays} into the dynamics (as is unavoidable in a cosmological context).  We show that, in such models, the 
infinite tower of initial conditions $\partial_t^{(n)}\phi$ at $t=0$ can be swapped for the freedom to fix the time evolution over some interval $\Delta t = T$ 
by requiring 
\begin{equation}
\label{IFF}
  \phi(t) = \phi_0(t) \hspace{5mm}\mathrm{on}\hspace{5mm}t \in \left[-T,0\right]
\end{equation}
with $\phi_0(t)$ the ``initial function'' and $T^{-1}$ related to the energy scale at which nonlocal effects become important.  This new formulation has 
considerable intuitive appeal since it makes manifest the nonlocal structure of the theory; the Cauchy surface has been ``smeared out'' over a
finite interval set by the underlying scale of nonlocality.

This mathematical re-formulation of the initial value problem suggests a rather novel new perspective on the key problems of predictivity and stability discussed above.  Although
the initial function is arbitrary from a purely mathematical viewpoint, we point out that $\phi_0(t)$ must be subjected to some auxiliary constraints in order
to ensure that the $-T \leq t \leq 0$ evolution is physically sensible.  For example, in the case of D-brane decay $\phi_0(t)$ should be restricted to the 
physically meaningful region of the potential.  The physical interpretation of these constraints is completely obfuscated in the ``old'' formulation of the 
initial value problem wherein one fixes infinitely many derivatives at $t=0$.

Remarkably, these mild physical restrictions on $\phi_0(t)$ have the effect of constraining the theory to a stable subset of
its solution space.  Although the Hamiltonian is globally unbounded below, the unstable directions in phase space cannot be accessed
from an allowed initial configuration.  This restriction is nonperturbative and retains information about the full nonlocal structure of the equations.
Moreover, the constrained theory makes robust predictions for the macroscopic ($\Delta t \gg T$) dynamics. 

In light of these results, we argue that there exists a broad class of stable, predictive nonlocal models whose equations of motion admit infinitely many
initial data.  A specific example is provided by D-brane decay in the background of a linear dilaton \cite{simeon,BMNR,delays}. Previous examples
of unstable solutions in this model were implicitly assuming unphysical initial conditions.  This perspective on the nature of the instability was not 
appreciated in previous studies because the initial value problem was formulated in a highly unnatural way.

Our results help to establish the theoretical consistency of nonlocal cosmological models, and 
also provides a plausible mechanism by which the consistent inclusion of closed strings might enable SFT to evade the Ostrogradski instability.  
(This has been conjectured as a possible solution of the instability problems that plague SFT, however, to our knowledge no concrete mechanism
has previously been proposed.)  Since we do not rely on any intrinsically stringy properties of the theory, it will be straightforward to apply our
approach to more general nonlocal models, in cosmology or elsewhere.  At the practical level, our analysis provides a new tool-box of efficient and 
numerically stable techniques for solving nonlinear equations with infinitely many derivatives.

The organization of this paper is as follows.  In section \ref{sec:delays} we discuss some general properties of delayed differential equations which will 
be important for the subsequent analysis.  In section \ref{sec:string} we re-visit models of light-like D-brane decay in light of the formulation (\ref{IFF}) of 
the initial value problem, showing how macroscopic predictivity and stability can be salvaged.  In section \ref{sec:QFT} we show how the simple-minded 
inclusion of dissipation can lead to delay-type nonlocality in a very general nonlocal QFT framework and we show how the formulation (\ref{IFF}) of the 
initial value problem can be implemented.  In section \ref{sec:timelike} we study a particular model, cosmological D-brane decay, showing again how 
stability/predictivity can be salvaged.  Finally, in section \ref{sec:conclusions}, we conclude.

\section{Nonlocal Models with Delays}
\label{sec:delays}

In this section we will provide a brief discussion of some general features of models whose nonlocality arises purely through delay operators of the form
$e^{-T\partial_t}$.  As we will see shortly, this type of nonlocality is often associated with the presence of some source of dissipation in the dynamics
(such as cosmological friction).

\subsection{The Method of Steps}
Consider a 
prototype model whose dynamics are governed by the following delay differential equation
\begin{equation}
\label{dde}
  \sum_{n=0}^{N} a_n \partial_t^{(n)} \phi(t) = F\left[ \phi(t-T) \right]
\end{equation}
Here $F\left[x\right]$ is some nonlinear forcing term and we assume that $T > 0$.  Equation (\ref{dde}) is a nonlinear differential equation of infinite order,
which is easy to see by noting the identity $\phi(t-T) = e^{-T\partial_T} \phi(t) = \sum_n \frac{(-T)^n}{n!} \partial_t^{(n)}\phi(t)$.  Infinite order equations have been
studied in some detail using the formal generatrix calculus in \cite{niky1} and also \cite{nl_rev,niky2}.  (See
\cite{gomis,woodard2,woodard3,calcagni} for different approaches.)

\begin{figure}[htbp]
\centerline{{\epsfxsize=0.5\textwidth\epsfbox{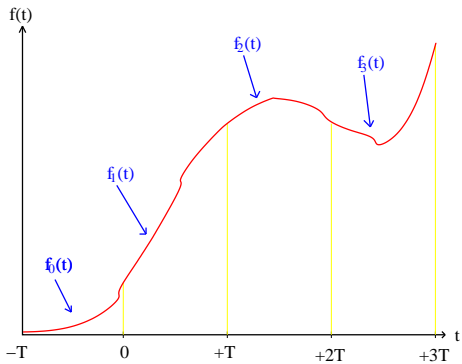}} }
\caption{Schematic illustration of the procedure for splitting a function up into solution segments defined over contiguous intervals of length $\Delta t = T$.
}
\label{fig:segments}
\end{figure}

The model (\ref{dde}) belongs to a special class of nonlocal theories, referred to as having ``compact support'' in \cite{woodard}, which can be formulated in a surprisingly simple
and intuitive manner using the method of steps (see also \cite{delays}).  The key observation is that equation (\ref{dde}) defines a mapping from functions on the interval $\left[t-T,t\right]$ to the 
interval $\left[t,t+T\right]$.  We can therefore break the solution $\phi(t)$ into segments $\phi_i(t)$ defined over contiguous intervals of length $T$.  Explicitly, we set
\begin{equation}
  \phi(t) = \phi_i(t)\hspace{5mm}\mathrm{on}\hspace{5mm}t\in\left[(i-1)T, i T\right]
\end{equation}
where the integer $i=0,1,2,\cdots$ labels the segment under consideration.  This procedure for splitting a solution up into segments is illustrated schematically in Fig.~\ref{fig:segments}.
The dynamical equation (\ref{dde}) provides a recursion relation to obtain the $i$-th solution segment:
\begin{equation}
\label{step}
  \sum_{n=0}^N a_n \partial_t^{(n)} \phi_i(t) =  F\left[\phi_{i-1}(t-T)\right], \hspace{5mm}i=1,2,\cdots
\end{equation}
On the first segment ($i=0$) we have an arbitrary ``initial function'' $\phi_0(t)$.  More on this shortly.

Starting from a given $\phi_0(t)$, the stepping procedure (\ref{step}) is very easy to implement.  Each solution segment is obtained by solving an 
inhomogeneous linear differential equation whose source term is constructed from the previous solution segment.  At the points $t=i T$ where the segments are joined we should
impose boundary conditions that $\phi(t)$ be continuous and as smooth as possible.  Since the differential operator on the left-hand-side of (\ref{step})
is $N$-th order, we can demand that
\begin{equation}
  \left.\partial_t^{(n)}\phi_{i+1}(t)\right|_{t=iT} = \left.\partial_t^{(n)}\phi_{i}(t)\right|_{t=iT}
  \hspace{5mm}\mathrm{for}\hspace{5mm} n = 0, \cdots, N-1
\end{equation}
This matching should be performed at each knot: $i=0,1,\cdots$  The method of steps is nonperturbative and solutions obtained using the recursion (\ref{step}) are often much more
efficient and numerically stable than those obtained using alternative approaches (such as higher order perturbation theory \cite{niky1} or the diffusion equation technique \cite{mulryne}).

Notice that, in general, the $N$-th derivative of $\phi(t)$ will be discontinuous at $t=0$.  This discontinuity will propagate into higher derivatives in subsequent segments.
The only way to avoid this effect is by choosing $\phi_0(t)$ to be a solution of (\ref{dde}).  In practice, however, we can minimize derivative discontinuities by using perturbation 
theory to construct the initial function.  None of our qualitative results in this work will rely on the smoothness properties of the solutions in any way.

\subsection{The Initial Value Problem}

The initial value problem associated with (\ref{dde}) can be formulated as
\begin{equation}
\label{IFF2}
  \phi(t) = \phi_0(t) \hspace{5mm}\mathrm{on}\hspace{5mm}t\in\left[-T,0\right]
\end{equation}
where the initial function $\phi_0(t)$ specifies the state of the system on the first segment.  Equation (\ref{IFF2}) can be regarded as 
``smearing out'' the Cauchy surface over a finite width.  Notice that (\ref{IFF2}) is not the unique formulation of the initial value problem for this model.  
If we have adopted perturbation theory (instead of the method of steps) it would have been most natural to specify the derivatives 
$\partial_t^{(n)}\phi$ at $t=0$, following \cite{niky1}.  Although both views of the initial value problem are mathematically valid, the formulation (\ref{IFF2}) 
is much more natural here because it makes the manifest the delay-type nonlocality of the underlying theory.

The initial value problem (\ref{IFF2}) is unusual.  The conventional notion of causality\footnote{For our purposes, the conventional notion of causality 
can be caricatured as ``the present determines the future''.} is violated, because the evolution depends on past history.  Suppose 
that we are observing the dynamics of $\phi(t)$ over a time scale short compared to the delay, $\Delta t \ll T$.  Equations (\ref{dde}) and (\ref{IFF2}) imply that the system 
will be \emph{directly} influenced by events which occurred very far in the past, before the start of the ``experiment''.  No finite amount of information about the state of 
the system at $t=0$ is sufficient to uniquely determine the $t>0$ evolution.  Indeed, equation (\ref{IFF2}) means that we could arrange for
\emph{any} time dependence over an interval $\Delta t = T$, regardless of the specific details of the dynamical equation (\ref{dde}).  Clearly, the nonlocality of the system is
associated with a breakdown of predictivity over \emph{microscopic} scales, $\Delta t \ll T$.

On the other hand, if we observe the system for a \emph{macroscopic} time scale, $\Delta t \gg T$, then the situation is different.  We retain a weakened version of causality
because the ``memory'' is small as compared to the dynamical scales of interest.  We could hope to measure the evolution over an entire segment and use this as an initial 
function to robustly predict the subsequent dynamics.  (In principle this still requires making infinitely many measurements and, in a realistic model, it 
might only be possible to reconstruct the initial function to some limited accuracy.  Any uncertainty in $\phi_0(t)$ propagates into all future
predictions.  In practice, this effect is usually tiny because the late-time dynamics are insensitive to the details of the initial function.)

Notice that, although the initial function may be ambiguous in some circumstances, it is \emph{not} completely arbitrary.  We should subject $\phi_0(t)$ to some auxiliary
constraints in order to ensure that the $-T\leq t \leq 0$ evolution was sensible.  The specific nature of these constraints will, obviously, depend on the system that 
we are modeling.  Notice that we have \emph{already} seen an example of this necessity: arbitrary choices of $\phi_0(t)$ will lead to derivative discontinuities.  We will
see shortly how the imposition of physical restrictions on $\phi_0(t)$ can have very significant implications for the $t>0$ dynamics of the system.

\section{Delayed Dynamics in String Theory}
\label{sec:string}

In the last section we introduced a general class of models with delay-type nonlocality, equation (\ref{dde}).  We argued that (\ref{IFF2}) provides the most natural formulation
of the initial value problem.  Let us now consider two explicit examples from string theory and show how this new perspective on the initial value problem provides invaluable insight
into the key problems -- stability and predictivity -- discussed in the introduction.

\subsection{An Example from $p$-adic String Theory}
\label{subsec:padic}

\begin{figure}[htbp]
\centerline{{\epsfxsize=0.5\textwidth\epsfbox{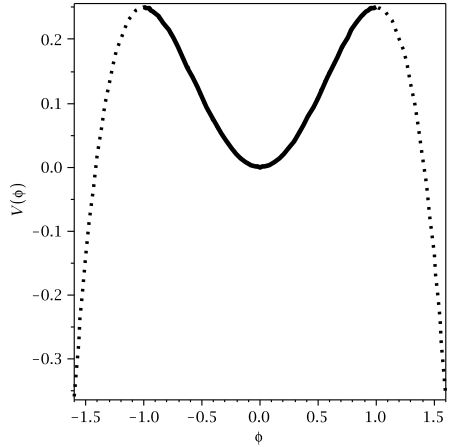}} }
\caption{The potential for the $p$-adic tachyon field $\phi$ taking $p=3$ for illustration.  The maxima
$\phi=\pm 1$ correspond to the physical state containing an unstable D-brane while the minimum at $\phi=0$ is the ``vacuum''
containing neither D-brane nor open string excitations.  The unbounded regions are thought to be associated with the closed string
tachyon.  The solid part of the potential, $-1 < \phi < +1$, denotes the range of field values which can be accessed by physically meaningful
initial functions.
}
\label{fig:padicpot}
\end{figure}

Let us now consider an explicit example of the type (\ref{dde}) whose nonlinear dynamics are particularly transparent.  Our example is based on $p$-adic string theory \cite{padic},
a toy model of the bosonic string which is known to reproduce a number of key features of the full string theory. (See \cite{thermal} for a review of some of 
these features and \cite{thermal,thermal2} for a discussion of the $p$-adic theory at finite temperature and implications for the cosmological constant problem.)  
The $p$-adic theory contains a single scalar field $\phi(x)$, the open string tachyon, that encodes the instability of a space-filling D-brane.  
Following \cite{simeon} we adopt light-cone coordinates $x^{\pm} = \frac{1}{\sqrt{2}}(x^0 \pm x^1)$ so that the metric takes the form $ds^2 = -(dx^0)^2 + (dx^1)^2 = -2dx^{+}dx^{-}$ and 
assume a linear dilaton profile: $\Phi(x) = -V^{+} x^{-}$.  In this closed string background the $p$-adic Lagrangian is \cite{simeon}
\begin{equation}
\label{padic}
  \mathcal{L} = \frac{e^{V^{+}x^{-}}}{g_p^2}\left[ -\frac{1}{2}\left( p^{-\frac{1}{2} \Box}\phi \right)^2 + \frac{1}{p+1} \phi^{p+1} \right]
\end{equation}
where $p$ is a prime number, $g_p$ is related to the open string coupling, we have set $\alpha' \equiv m_s^{-2} \equiv 1$.  The potential is
\begin{equation}
\label{padic_pot}
  g_p^2 e^{-V^{+} x^{-}}V(\phi) = +\frac{1}{2}\phi^2 - \frac{1}{p+1}\phi^{p+1}
\end{equation}
which is plotted in Fig.~\ref{fig:padicpot}.  If we assume a light-like tachyon profile $\phi=\phi(x^{+})$ then the equation of motion derived from 
(\ref{padic}) takes a remarkably simple form
\begin{equation}
\label{padic_KG}
  \phi(x^+) = \phi^p(x^+ - T)
\end{equation}
where the delay is
\begin{equation}
  T \equiv V^{+} \ln p
\end{equation}

Equation (\ref{padic_KG}) admits two constant solutions: $\phi=0$ and $\phi=1$ (also $\phi=-1$ when $p$ is odd) corresponding, respectively, to the local 
maxima and minima of the potential.  Physically, the solution $\phi=1$ corresponds to state with an unstable space-filling D-brane while $\phi=0$
is the empty state containing neither D-brane nor open string excitations.\footnote{This configuration is sometimes called the ``vacuum'' although,
properly, the theory (\ref{padic}) has no minimum energy state.}  The dynamical process of rolling the tachyon from the unstable maximum
to the minimum of the potential -- tachyon condensation -- provides a time-dependent description of D-brane decay in the $p$-adic string theory.

The initial value problem (\ref{IFF2}) requires that we choose the evolution of the $p$-adic tachyon over a time interval $\Delta x^+ = T = V^+ \ln p$.  
Which choices of $\phi_0(x^+)$ will lead to a sensible physical interpretation in terms of brane decay?  To answer this question, first notice that
the potential (\ref{padic_pot}) is unbounded below for $\phi > 1$ (also $\phi < -1$ for $p$ odd).  This is usually thought to be associated with the 
presence of the closed string tachyon \cite{unbounded} and solutions rolling down the unbounded region are regarded as unphysical.  Second,
recall that near $\phi=0$ there are no open string excitations (because there is no brane for them to end on) and hence the physical meaning of the 
open string variable $\phi$ is obscure.  We should therefore expect that the theory (\ref{padic}) only admits a robust physical interpretation in terms of 
brane decay when the initial function satisfies:
\begin{equation}
\label{padic_constraint}
  0 < \phi_0(x^+) < 1
\end{equation}
When $p$ is odd we can continue this to $-1 < \phi_0(x^+) < 0$.  Thus, only the solid region of the potential curve plotted in Fig.~\ref{fig:padicpot} 
can be accessed by physically meaningful initial functions.

Equation (\ref{padic_KG}) can be easily solved using the method of steps.  However, it happens that we can write down a simple exact
solution which is valid on \emph{all} segments,  as was pointed out in \cite{BMNR}.  That solution is
\begin{equation}
\label{padic_soln}
  \phi(x^+) = \exp\left[ -e^{x^+/ V^+} F(x^+) \right]
\end{equation}
where $F(x^+)$ is a periodic function satisfying 
\begin{equation}
\label{F_periodic}
  F(x^+) = F(x^+-T)
\end{equation}
(The solution (\ref{padic_soln}) is always positive, however, when $p$
is odd we can obtain a second solution simply by flipping the sign.)  The initial value problem (\ref{IFF2}) fixes the free function $F(x^+)$ 
on the first segment as
\begin{equation}
\label{F}
  F(x^+) = -e^{-x^+ / V^+} \ln \left[ \phi_0(x^+) \right]  \hspace{5mm}\mathrm{on}\hspace{5mm}t\in\left[-T,0\right]
\end{equation}
The periodicity relation (\ref{F_periodic}) then uniquely determines $F(x^+)$ on all subsequent segments.

\begin{figure}[htbp]
\centerline{{\epsfxsize=0.5\textwidth\epsfbox{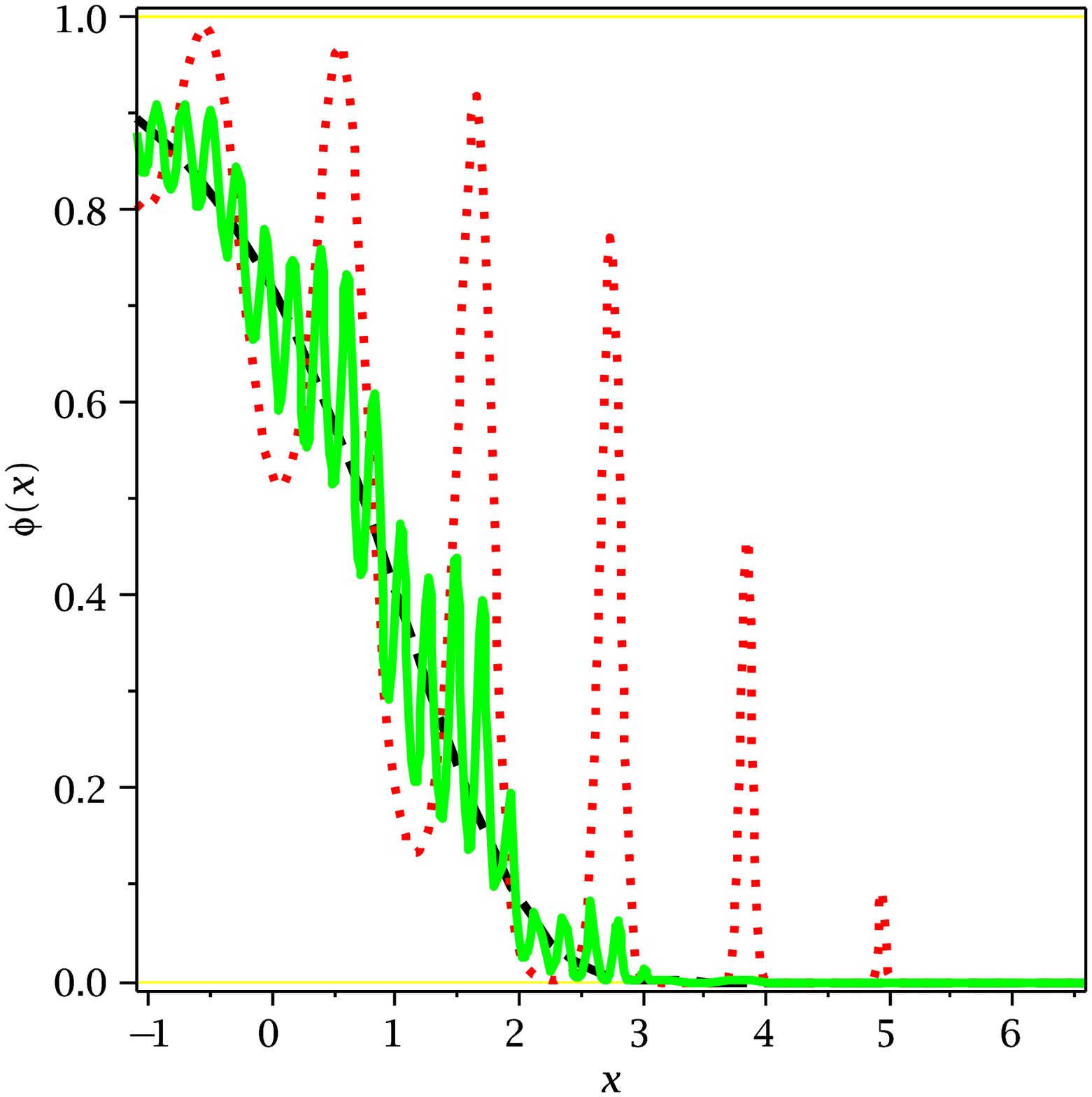}}
{\epsfxsize=0.5\textwidth\epsfbox{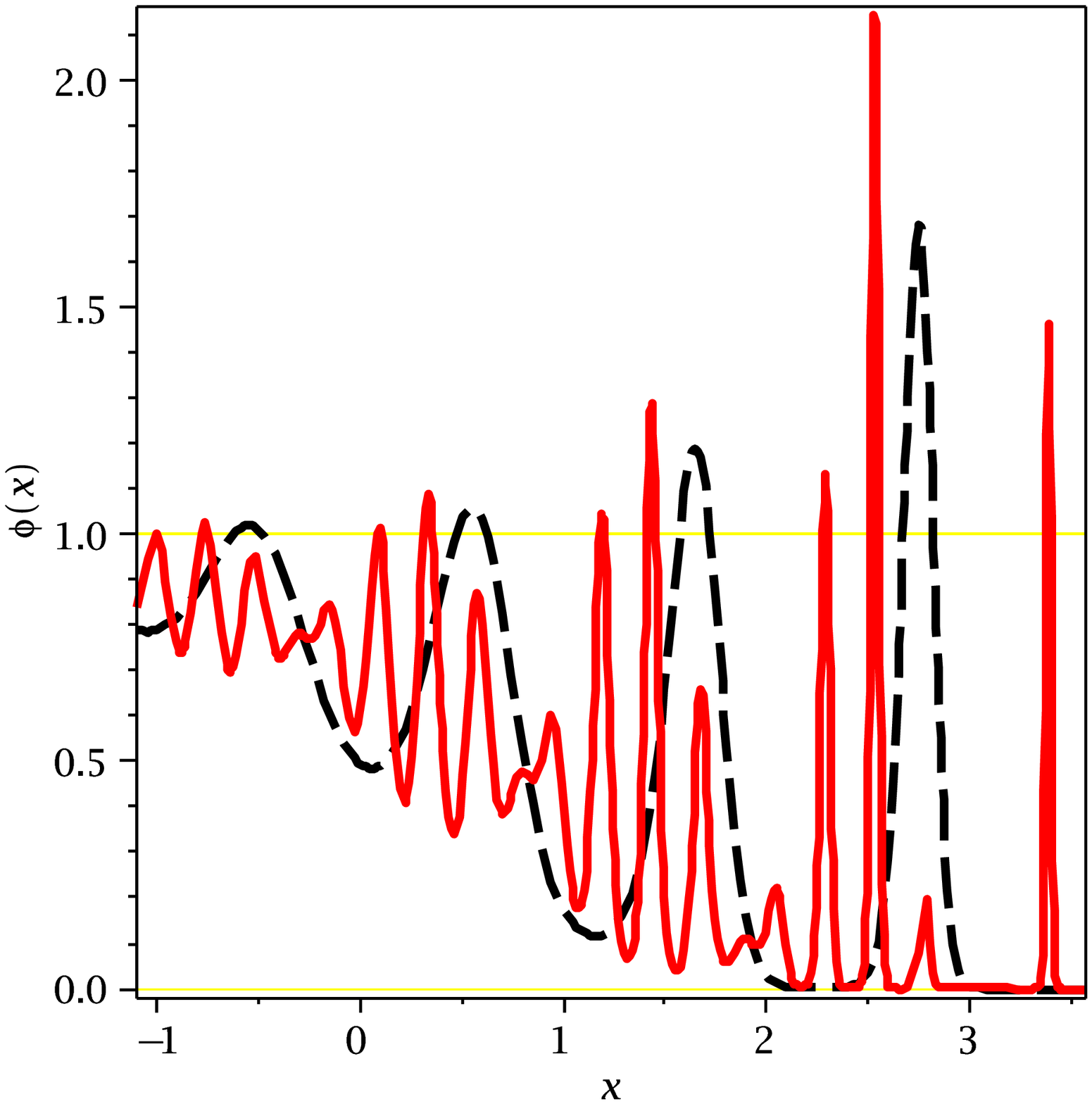}}}
\caption{
The time evolution of the $p$-adic tachyon field for several representative choices of the initial function.  In the left panel we consider only
examples which satisfy the constraint $0 < \phi_0(x^+) < 1$.  At late times the field settles down to the minimum of the potential
and the macroscopic ($\Delta x^+ \gg T$) dynamics are largely insensitive to the details of $\phi_0(x^+)$.  In the right panel we consider two examples
where the initial function \emph{violates} the physical constraint $0 < \phi_0(x^+)<1$.  Now the field undergoes erratic oscillatory behaviour at late 
times and the dynamics become highly sensitive to the details of $\phi_0(x^+)$.  In both panels we set $p=3$ and $V^+=1$ for illustration.  The horizontal
yellow lines illustrate the critical values $\phi=0$ and $\phi=1$.
}
\label{fig:padic}
\end{figure}

Armed with the exact closed-form solution (\ref{padic_soln}) we can address the nonlinear dynamics of (\ref{padic}) is a remarkably simple way.  If 
the initial function satisfies (\ref{padic_constraint}) then we must have $F(x^+)>0$ and the solution (\ref{padic_soln}) approaches the minimum 
$\phi=0$ very rapidly for $t \gsim T$.  This behaviour is illustrated in the left panel of Fig.~\ref{fig:padic}.  Notice
the peculiar dynamics: $\phi(x^+)$ can roll to higher potential energy during the rolling, even though the dynamics are governed 
by gradient flow.  Such motion is possible because the kinetic energy in this theory can be negative.  Nevertheless, the 
nonlinear dynamics are stable in the sense that the field \emph{does} eventually settle down to the minimum of the potential at late times.  Note also
that these late-time asymptotics are largely insensitive to details of $\phi_0(x^+)$.

The Hamiltonian associated with (\ref{padic}) is unbounded from below \cite{BMNR}.  However, we have just argued that the nonlinear solutions of this 
theory do \emph{not} display any unstable behaviour.  How can this be?  To understand what is going on, we should imagine relaxing the constraint 
(\ref{padic_constraint}) on the initial function.  Now the dynamics associated with (\ref{padic_KG}) are \emph{very} different.  If $\phi_0(x^+) > 1$ 
anywhere on $x^+ \in \left[-T,0\right]$ then we must have $F(x^+) < 0$ at one or more points and consequently the solution (\ref{padic_soln}) undergoes 
wild oscillations with divergent amplitude.  This is a manifestation of the Ostrogradski instability.  These dynamics are illustrated in the right panel of 
Fig.~\ref{fig:padic}.  As can be seen, the qualitative late-time behaviour is now \emph{highly} sensitive to the details of $\phi_0(x^+)$.

In the formulation (\ref{IFF2}) it is quite apparent what is going wrong with the solutions plotted in the right panel of Fig.~\ref{fig:padic}.  We have allowed the tachyon to
sample unphysical regions of the potential on the initial segment $-T \leq t \leq 0$.  However, this  conclusion would be completely obfuscated if we 
adopted a perturbative approach.  Writing $\phi(x^+) = 1 + \delta\phi(x^+)$ we obtain
\begin{equation}
\label{padic_pert}
  \left[ p^{V^+ \partial_+} - p \right] \delta \phi(x^+) = 0
\end{equation}
to linear order in $\delta \phi$.  The solution of (\ref{padic_pert}) is
\begin{equation}
\label{padic_pert_soln}
  \delta \phi (x^+) = \sum_{n=-\infty}^{+\infty} \alpha_n e^{M_n x^+}
\end{equation}
which depends on infinitely many free parameters $\alpha_n$.  The frequencies $M_n$ correspond to the zeroes of the generatix $f(s) = p^{V^+ s} - p$.  
Explicitly, we have
\begin{equation}
\label{Mn_padic}
  M_n = \frac{1}{V^+}\left[1 + \frac{2 \pi i n}{\ln p}\right], \hspace{5mm} n= 0, \pm 1, \pm 2, \cdots
\end{equation}
The spectrum (\ref{Mn_padic}) can be interpreted as a single growing mode (the usual tachyon) along with an infinite tower of modes which oscillate
with ever-growing amplitude.\footnote{The solution (\ref{padic_pert_soln}) is completely consistent with (\ref{padic_soln}).  This is easily seen by 
expanding $F(x^+)$ in a Fourier series as 
\begin{equation}
\label{F_fourier}
  F(x^+) = a_0 + \sum_{n=1}^{\infty} a_n \cos\left(\frac{2\pi n x^+}{ T }\right) + \sum_{n=1}^{\infty} b_n \sin\left(\frac{2\pi n x^+}{ T }\right) 
\end{equation}
When $1 \gg |a_n|, |b_n|$ we can expand (\ref{padic_soln}) in a Taylor series and trivially recover $\phi(x^+) \cong 1 + \sum_n \alpha_n e^{M_n x^+}$.}
Half of those spurious modes carry wrong-sign kinetic energy and represent ghost-like excitations \cite{BMNR}.
In \cite{BMNR} it was noted that there exists an ``island of stability'' in initial condition space.  By carefully tuning the coefficients $\alpha_n$ in the 
solution (\ref{padic_pert_soln}) we can have sensible late-time dynamics.  This tuning looks highly contrived in the perturbative approach, however, in 
light of the method of steps we see that it amounts to simply satisfying the physical bound (\ref{padic_constraint}) over a microscopic time 
$\Delta x^+ = T$.

\begin{figure}[htbp]
\centerline{{\epsfxsize=0.5\textwidth\epsfbox{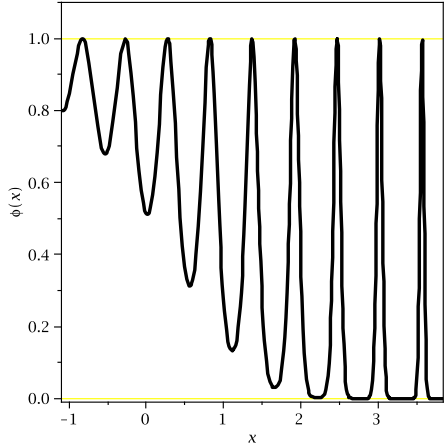}}}
\caption{
The time evolution of the $p$-adic tachyon field for the special case where the initial function crosses $\phi_0(x^+) = 1$ at some
point on $x^+\in\left[-T,0\right]$.  In this special case the tachyon undergoes sustained periodic oscillations between the minimum and 
maximum of the potential.  It may be possible  to interpret this solution as an array of space-like branes (S-branes).  We set $p=3$ and 
$V^+=1$ for illustration.  The horizontal yellow lines illustrate the critical values $\phi=0$ and $\phi=1$.
}
\label{fig:sbrane}
\end{figure}

Before moving on, it is worth commenting on a special class of initial functions which we have excluded from our discussion up to now.
In the special case where $\phi_0(x^+)=1$ for some points on $x^+ \in \left[-T,0\right]$, then sustained oscillations are possible.  At late times
the tachyon interpolates periodically between the maximum and minimum, as is illustrated in Fig.~\ref{fig:sbrane}.  Taken at face value, this solution
describes a brane which repeatedly dissolves into tachyon matter and then re-assembles itself.  We are tempted to interpret this
as an array of space-like brane (S-brane) \cite{sbrane} in the $p$-adic string theory.  For the time being, however, we remain agnostic as to whether the 
upper limit $\phi_0(x^+) = 1$ should be included in the physical constraint (\ref{padic_constraint}).

The oscillatory behaviour in Fig.~\ref{fig:sbrane} might seem very surprising.  The dynamics of the theory (\ref{padic}) is purely dissipative.  
How can this be reconciled with the observation that
solutions need not limit to $\phi=0$ at late times?  The key to this puzzle is noting that the dilaton profile $\Phi(x)$ is nearly constant on very small 
scales $\Delta x^+ \ll T \sim V^{+}$.  Deep in the UV the dissipative effects associated with the dilaton gradient must be negligible.  This observation 
is related to the fact that the spectrum (\ref{Mn_padic}) contains frequencies $M_n$ which are arbitrarily large as compared to the energy scale 
associated with the damping, $T^{-1}$.

\subsection{An Example from String Field Theory}
\label{subsec:sft}

\begin{figure}[htbp]
\centerline{{\epsfxsize=0.5\textwidth\epsfbox{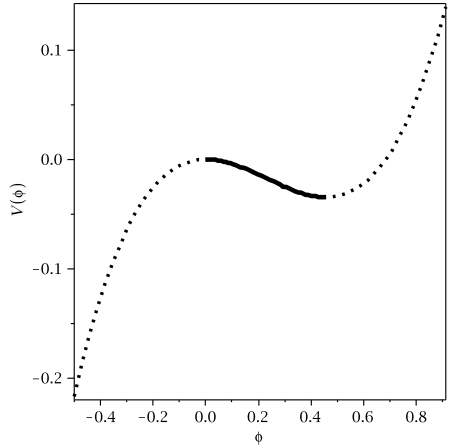}} }
\caption{The potential for the SFT tachyon field.  The maximum $\phi=0$ correspond to the physical state 
containing an unstable D-brane while the minimum at $\phi=C^{-3}$ is the ``vacuum'' containing neither D-brane nor open string excitations.  
The unbounded region is thought to be associated with the closed string tachyon.  The solid part of the potential, $0 < \phi < C^{-3}$, denotes 
the region which can be accessed by physically sensible initial functions.
}
\label{fig:SFTpot}
\end{figure}

In this subsection we consider a somewhat more realistic example: the level zero truncation of bosonic SFT.  In order to obtain a delayed formulation
we again assume a light-like dilaton profile of the form $\Phi(x) = - V^{+}x^{-}$.  The Lagrangian is \cite{simeon}
\begin{equation}
\label{light}
  \mathcal{L} = \frac{e^{V^{+}x^{-}}}{g_s^2}\left[ - \frac{1}{2} (\partial\phi)^2 + \frac{1}{2}\phi^2 - \frac{C^{3}}{3}\left(C^{+\Box}\phi\right)^3 \right]
\end{equation}
where $g_s$ is the open string coupling, the dimensionless constant\footnote{Our $C$ is related to the constant $K$ in employed by Hellerman and 
Schnabl in \cite{simeon} as $C=1/K^{\mathrm{H.S.}}$ and to the constant $K$ employed by Beaujean and Moeller in \cite{delays} as $C=K^{\mathrm{B.M.}}$.} is 
$C = 3\sqrt{3} / 4 \cong 1.23$  and the scalar field $\phi(x)$, again, corresponds to the open string tachyon.  
The potential is
\begin{equation}
\label{sft_pot}
  g_s^2 e^{-V^{+}x^{-}} V(\phi) = -\frac{1}{2}\phi^2 + \frac{C^3}{3}\phi^3
\end{equation}
which is plotted in Fig.~\ref{fig:SFTpot}.  Assuming a light-like profile $\phi=\phi(x^+)$ the equation of motion takes the form
\begin{equation}
\label{light_KG}
  \left(V^{+}\partial_+ - 1\right)\phi(x^+) + C^3 \phi^2(x^+ - T) = 0
\end{equation}
where the delay is
\begin{equation}
  T = 2 V^{+} \ln C
\end{equation}

Equation (\ref{light_KG}) cannot be solved analytically.  However, the method of steps is easy to implement numerically.  
As in the $p$-adic case, we should subject 
$\phi_0(x^+)$ to some restrictions to ensure that the solutions admit a meaningful physical interpretation in terms of D-brane decay.  
What class of initial functions are allowed?  As in our previous example, the maximum $\phi=0$ corresponds to the physical configuration with an 
unstable space-filling D-brane while the minimum $\phi=C^{-3}$ corresponds to the empty configuration which contains neither D-brane nor open 
string excitations.  At the minimum, the meaning of the open string variable $\phi$ is obscure.  Similarly, solutions rolling down the unbounded 
$\phi < 0$ side of the potential are associated with the closed string tachyon and considered unphysical \cite{unbounded}.  All told, the theory 
(\ref{light}) admits a robust physical interpretation in terms of D-brane decay only when the initial function satisfies
\begin{equation}
\label{sft_constraint}
  0 < \phi_0(x^+) < C^{-3}
\end{equation}
which corresponds to the solid region of the potential curve plotted in Fig.~\ref{fig:SFTpot}.  In addition to (\ref{sft_constraint}), we should also 
ensure that the tachyon rolls down the ``correct'' side of the potential, towards the open string vacuum at $\phi=C^{-3}$.  We will discuss this point 
in more detail shortly, however, a sufficient (but not necessary) condition is the following
\begin{equation}
\label{sft_constraint2}
  \phi_0(0) > \phi_0(-T)
\end{equation}
The condition (\ref{sft_constraint2}) allows for a variety of non-monotonic behaviours on the interval $x^+\in \left[-T,0\right]$, however, it imposes that
the ``net'' motion of the tachyon over this time is to increase.  As was the case with (\ref{sft_constraint}), the condition (\ref{sft_constraint2}) is
not onerous: one expects it to be satisfied for any solution that admits a physical interpretation in terms of D-brane decay.


We have studied the nonlinear solutions obtained 
using the method of steps and found that the dynamics agree very well with the results of \cite{BMNR} using the diffusion equation approach.  As long as the 
initial function $\phi_0(x^+)$ satisfies the physical constraints (\ref{sft_constraint},\ref{sft_constraint2}) then all solutions will eventually settle down 
to the minimum of the potential in a time comparable to the delay.   This behaviour is illustrated in the left panel of Fig.~\ref{fig:sft} for three representative
choices of initial function (provided explicitly in Appendix A).  It should be emphasized that our numerical evidence for this result is much more exhausitive than
the three specific examples plotted in Fig.~\ref{fig:sft}: we have repeated this analysis for a very broad array of initial functions and found qualitatively similar
behaviour in all cases.

The Hamiltonian associated with (\ref{light}) is unbounded below, however, the nonlinear solutions do not display any unstable behaviour as long as (\ref{sft_constraint}) is satisfied.  
Notice also that the late-time dynamics are insensitive to the details of $\phi_0(x^+)$.  

\begin{figure}[htbp]
\centerline{{\epsfxsize=0.5\textwidth\epsfbox{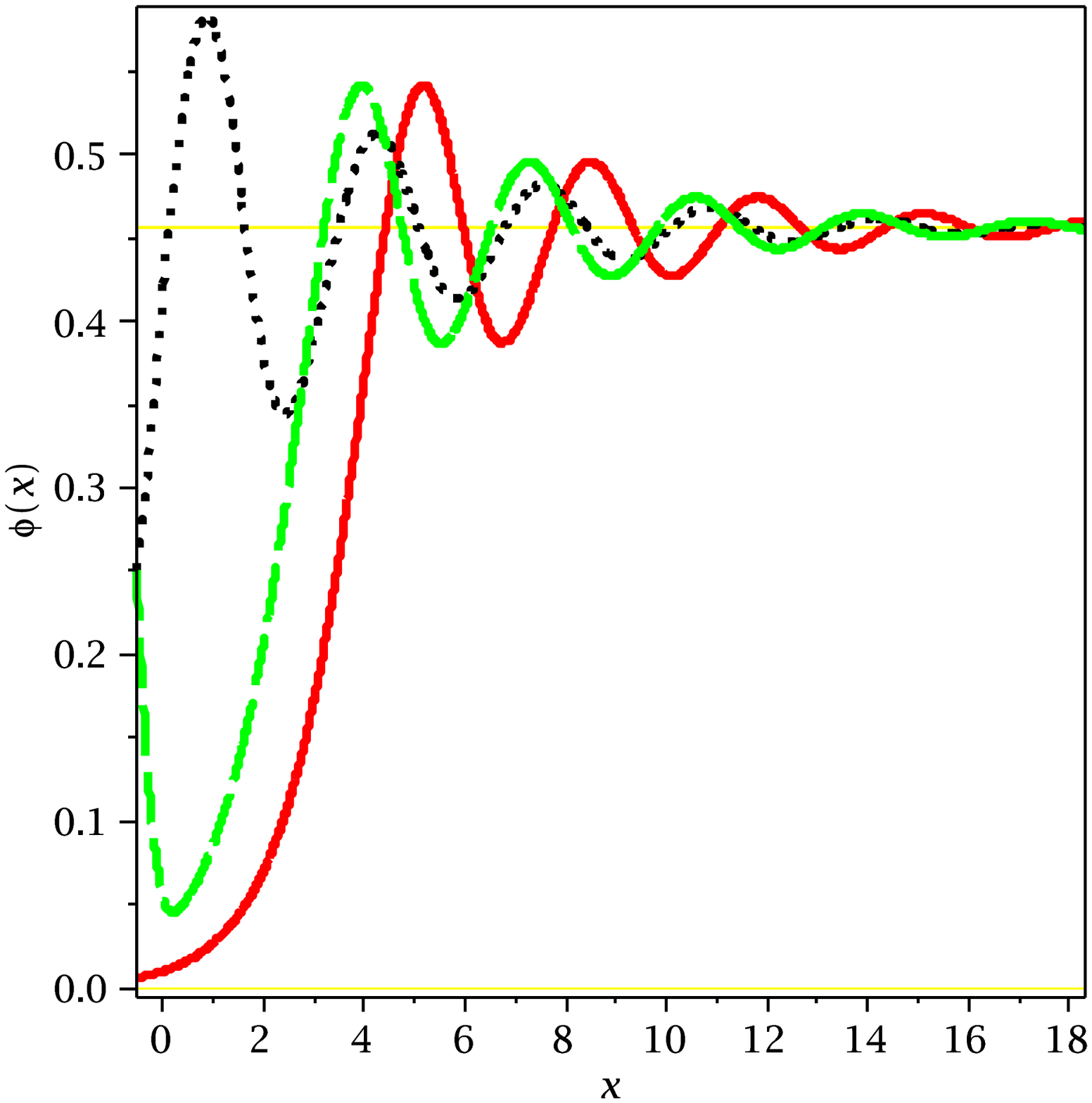}}
{\epsfxsize=0.5\textwidth\epsfbox{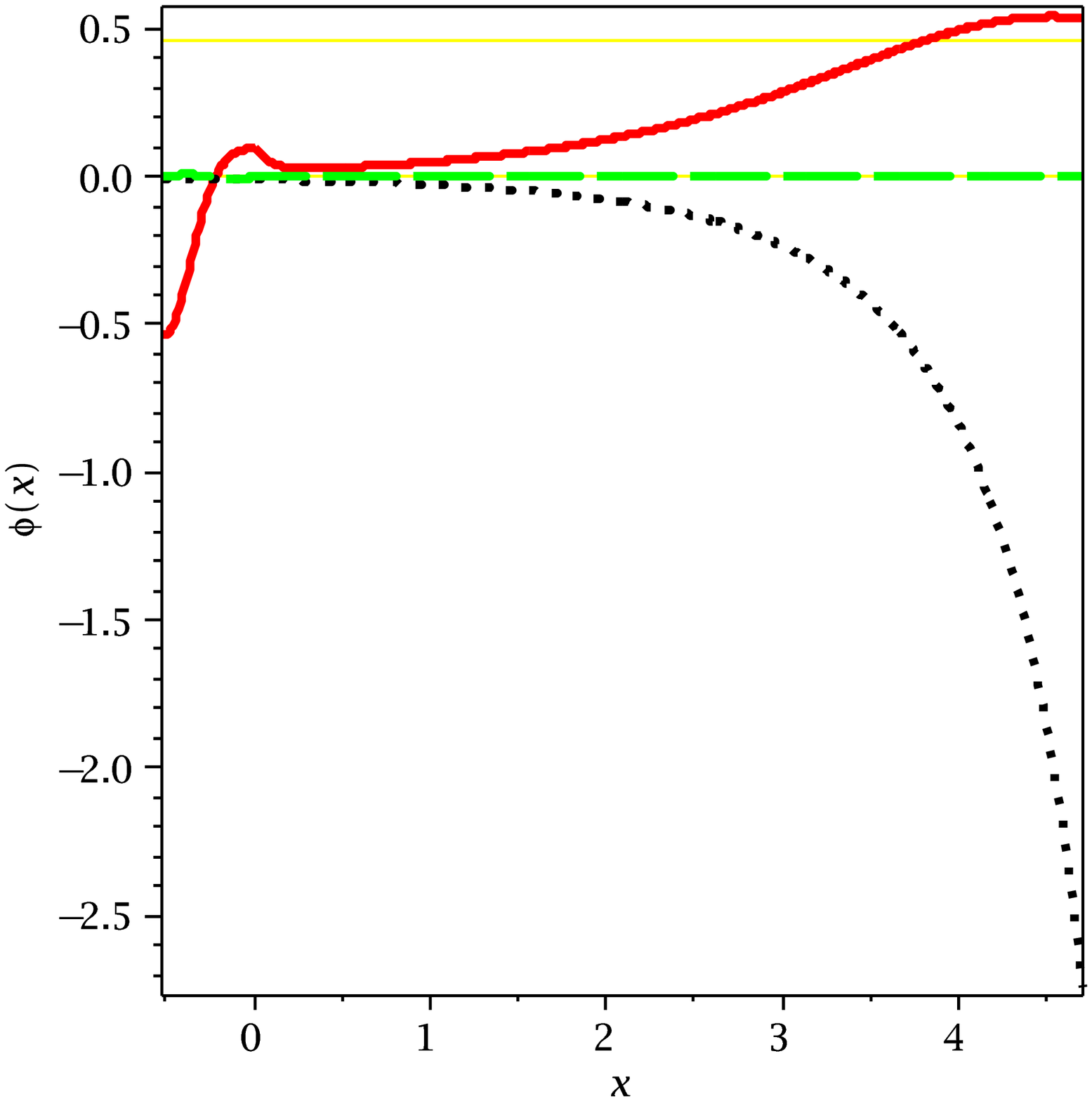}}}
\caption{
The time evolution of the light-like tachyon field for several representative choices of the initial function.  In the left panel we consider only
examples which satisfy the constraint $0 < \phi_0(x^+) < C^{-3}$.  At late times the field settles down to the minimum of the potential
and the macroscopic ($\Delta x^+ \gg T$) dynamics are largely insensitive to the details of $\phi_0(x^+)$.  In the right panel we consider three examples
where the initial function \emph{violates} the physical constraint $\phi_0(x^+) > 0$.  The solutions may grow without bound and the late-time 
dynamics are highly sensitive to the details of $\phi_0(x^+)$.  In both panels we have set $V^+=1$ for illustration and the horizontal yellow lines 
illustrate the critical values $\phi=0$ and $\phi=C^{-3}$.  The explicit choices of $\phi_0(x^+)$, for both panels, are provided in Appendix A.
}
\label{fig:sft}
\end{figure}

Just like in our last example, the dynamics are very different if we relax the physical constraint (\ref{sft_constraint}) on the initial function.
If it happens that $\phi_0(x^+) < 0$ on $x^{+}\in \left[-T,0\right]$ then the possibility arises for $\phi(x^+)$ to roll down the unbounded part of the potential
and grow limitlessly.\footnote{Interestingly, the constraint (\ref{sft_constraint}) is sufficient but not necessary to have sensible evolution at late
times.}  In this case the late-time dynamics will depend very sensitively on the details of $\phi_0(x^+)$.  This is illustrated in the right panel of Fig.~\ref{fig:sft}
for three representative choices of $\phi_0(x^+)$ (provided explicitly in Appendix A).

Because we lack an exact analytic solution of (\ref{light_KG}), it is difficult to prove rigorously that all solutions will relax to the minimum, once 
(\ref{sft_constraint},\ref{sft_constraint2}) are satisfied.  We have provided numerical evidence of this fact.  To reinforce our conclusions, we now provide 
also analytical arguments.  Firstly, we note that as soon as $\phi(x^+)$ approaches the basin of the closed string vacuum, $\phi=C^{-3}$, then its fate is sealed.
All solutions in the vicinity of this point undergo damped oscillations and eventually converge to the minimum, which is easily verified by perturbation theory.  Writing
\begin{equation}
  \phi(x^+) = C^{-3} + \delta \phi(x^+)
\end{equation}
and linearizing (\ref{light_KG}) in $\delta\phi$ we obtain
\begin{equation} 
\label{pert_new}
  \left[ V^{+}\partial_+ - 1 + 2 C^{-2V^{+}\partial_+}\right]\delta\phi = 0
\end{equation}
The solutions of (\ref{pert_new}) take the form
\begin{equation}
\label{sum_new}
  \delta\phi = \sum_n a_n e^{s_n x^+}
\end{equation}
where $a_n$ are arbitrary complex numbers and
\begin{equation}
  V^{+} s_n = 1 + \frac{1}{2 \ln C} W_n\left[ -\frac{4\ln C}{C^2} \right]
\end{equation}
where $W_n$ are the branches of the Lambert-W function and $n$ runs over all integer values.  One may verify that the $s_n$ are complex numbers which appear in conjugate 
pairs and also that $\mathrm{Re}\left[s_n\right] < 0$ for all $n$.  Thus, all solutions which approach $\phi=C^{-3}$ must eventually converge to the vacuum.

From the preceeding paragraph, we expect that any unbounded solutions must be associated with dynamics near the maximum $\phi=0$.  For example, one could imagine starting 
very close to $\phi=0$ with a large velocity that would allow the tachyon to roll over the maximum and down the unbounded side of the potential.  We have claimed that the
condition (\ref{sft_constraint2}) is sufficient to forbid such behaviour (and verified this claim also numerically).  Let us now seek to understand better the origin of this
constraint.  Consider an initial function $\phi_0(x^+)$ such that $\phi_0(0)$ is very close to zero.  Intuitively, such a function represents the most dangerous possibility.
By continuity, because $\phi_0(0)$ is near the maximum, then so is $\phi_1(0)$.  By writing the equation of motion (\ref{light_KG}) in delayed form we can establish the following
result:
\begin{eqnarray}
  V^+ \left. \partial_+ \phi_1\right|_{x^+=0} &=& \phi_0(0) - C^3 \phi_0^2(-T) \\
                                              &>& \phi_0(0) - \phi_0(-T)
\end{eqnarray}
where the inequality on the second line follows from (\ref{sft_constraint}).  If we further assume (\ref{sft_constraint2}) then we have
 \begin{equation}
  \left. \partial_+ \phi_1\right|_{x^+=0} > 0
\end{equation}
so that the solution on the first know must start near the maximum $\phi=0$ and roll towards the true minimum $\phi=C^{-3}$.  We can describe the early stages of this motion using perturbation
theory.  Writing
\begin{equation}
  \phi = 0 + \delta\phi
\end{equation}
and linearizing (\ref{light_KG}) we find the growing solution
\begin{equation}
  \delta\phi = a_0 e^{x^+ / V^+}
\end{equation}

The preceeding discussion suggests that our results are rather general.  As long as (\ref{sft_constraint},\ref{sft_constraint2}) are satisfied one expects that all solutions will roll 
away from the maximum, experiencing exponential growth, until they reach the vicinity of $\phi=C^{-3}$, at which point damped oscillations take over and the solution converges to the
minimum.  Indeed, precisely this behaviour has been observed in our numerical analysis.  This discussion is also consistent with \cite{BMNR}.

Finally, let us comment on a ``priviledged'' class of initial functions.  If we wish to minimize derivative singularities (which are presumably unphysical), we should choose $\phi_0(x^+)$
to be an approximate solution of the equation (\ref{light_KG}).  For solutions rolling away from the maximum, we have a one-parameter family of such functions: $\phi_0(x^+) = \epsilon \, e^{x^+/V^+}$
where $\epsilon \ll 1$ quantifies the initial displacement.  With this choice, the condition (\ref{sft_constraint2}) becomes unnecessary, because the dynamical equation (\ref{light_KG})
forces solutions near the maximum to grow monotonically.  The possibility to roll down the unbounded side of the potential while satisfying (\ref{sft_constraint}) arises \emph{only}
for initial functions that are very far from being solutions of (\ref{light_KG}).  Hence, this possibility will typically also be associated with unphysical derivative singularities.

\subsection{Stability and Predictivity}

The dynamics illustrated in  Fig.~\ref{fig:padic} and Fig.~\ref{fig:sft} are quite remarkable.  In both cases the Hamiltonian is unbounded from below.  Nevertheless,
nonlinear solutions cannot access the unstable directions in phase space starting from an ``allowed'' initial function $\phi_0(x^+)$.  The constraint which must be 
imposed on $\phi_0(x^+)$ in order to obtain sensible late-time evolution is \emph{extremely} mild: it amounts to restricting the tachyon to the physically meaningful region
of the potential over an interval comparable to the microscopic scale of nonlocality.\footnote{In particularly, we do not require that the field sits at an unstable maximum 
for an infinite amount of time.  That would be incompatible with quantum mechanics and, presumably, also with macroscopic causality/predictivity.}  These mild physical 
restrictions on $\phi_0(x^+)$ have the effect of constraining the theory to a stable subset of its solution space.  That this might be possible has been suggested previously,
for example by Woodard and Eliezer in \cite{woodard}.  The approach employed here has the advantage of 
being completely nonperturbative and also physically well-motivated.  Moreover, our solutions depend on infinitely many free coefficients and, in this sense, retain 
information about the full nonlocal structure of the model.  (Hence, our approach is different from perturbative localization.)

Suppose that we had adopted a different view of the initial value problem: fixing infinitely many derivatives $\partial_t^{(n)}\phi$ at $t=0$, instead of the initial 
function formulation (\ref{IFF2}).  In this picture we would have uncovered an ``island of stability'' in initial condition space which leads to stable late-time dynamics.  
Constraining the theory to the island of stability requires a seemingly delicate and contrived tuning amongst the infinite tower of initial conditions.  However, this
tuning \emph{seems} contrived only because one is using a highly unnatural formulation of the initial value problem.  In light of the initial function formulation (\ref{IFF2}) we 
see that, in fact, the constraints imposed by late-time stability are very mild.  

What becomes of predictivity in our delayed models (\ref{padic_KG}) and (\ref{light_KG})?  As discussed in section \ref{sec:delays}, there is a breakdown of predictivity
on \emph{microscopic} scales, $\Delta x^+ \ll T \sim V^+$.  In essence, this occurs because our ignorance about the past history of the system makes the initial function 
$\phi_0(x^+)$ inherently ambiguous.  However, as we have seen, the late-time dynamics are largely insensitive to $\phi_0(x^+)$, provided it belongs to the class of allowed 
initial functions.  Hence, we are still able to recover predictivity on \emph{macroscopic} scales, $\Delta x^+ \gg T$.  

We view the breakdown of microscopic predictivity as a necessary feature of working with a fundamentally nonlocal theory.  Indeed, it seems quite natural: we should not 
expect to be able to unambiguously resolve dynamics on times comparable to the underlying scale of nonlocality.

Obviously, our conclusions about the nonperturbative stability of equations (\ref{padic_KG}) and (\ref{light_KG}) are purely classical.  One might
worry that instabilities re-appear at the loop level (see \cite{thermal2} for a related discussion) or that the tachyon field $\phi$ can tunnel to the unbounded
region of the potential, violating the physical restrictions (\ref{padic_constraint}) and (\ref{sft_constraint}).  We note, however, that in string theory the 
situation may be rather subtle.  It is not clear if the physical degree of freedom associated with the tachyon field even \emph{exists} on the unbounded
region of the potential, or at the local minimum.\footnote{At the local minimum the D-brane has disappeared and there can be no open string degrees
of freedom.  It may be that $\phi$ somehow encodes the closed string gas which is the outcome of the brane decay \cite{open_closed} or it may simply
be that the $\phi$ disappears from the theory.  The latter interpretation is similar to what happens to the inflaton field during brane/anti-brane 
inflation \cite{brane_inf}.
At the end of inflation, the mobile brane annihilates with its anti-brane and the inflaton field (corresponding to the inter-brane separation) no longer
exists as a physical state in the theory.}

To summarize: the string theory models (\ref{padic_KG}) and (\ref{light_KG}) are nonperturbatively stable and make unambiguous predictions over
time scales large compared to the delay, provided $\phi_0(t)$ belongs to a class of allowed initial function.  Obviously, our optimistic conclusions have 
relied heavily on the delayed form of the prototype model (\ref{dde}) and the formulation (\ref{IFF2}) of the initial value problem.  Only a very privileged 
class of nonlocal theories (those having compact support \cite{woodard}) admit such a formulation.  One might wonder how much of our analysis can be carried 
over to a more realistic QFT framework.  We will investigate this question in section \ref{sec:QFT}.

\section{The Initial Function Formulation Applied to Generic Nonlocal Field Theory Models}
\label{sec:QFT}

Our goal now is to determine the extent to which the optimistic conclusions of section \ref{sec:string} carry over to a general nonlocal 
QFT framework.  To this end, we first introduce the very general class of nonlocal field theories.  Next, we will show how the simple-minded inclusion of 
dissipation into the dynamics leads to delay-type nonlocality and an initial value problem of the form (\ref{IFF2}).

\subsection{Prototype Nonlocal Field Theory}

Consider the prototype nonlocal field theory 
\begin{equation}
\label{proto}
  \mathcal{L} = \frac{1}{2} \phi F(\Box) \phi - \Vint(\phi),  \hspace{5mm} \Vint(\phi) = \frac{g}{3!}\phi^3 + \frac{\lambda}{4!}\phi^4 + \cdots
\end{equation}
This model subsumes known examples from SFT \cite{sft_rev}, $p$-adic string theory \cite{padic} and also cosmology 
\cite{phantom,mulryne,lidsey,padic_inf,ng1,ng2}.  The full scalar potential is 
\begin{eqnarray}
  V(\phi) &=& -\left.\mathcal{L}\right|_{\Box=0} \nonumber \\
  &=& -\frac{F(0)}{2} \phi^2 + \Vint(\phi) \nonumber \\
  &\equiv& \frac{m^2}{2!}\phi^2 + \frac{g}{3!}\phi^3 + \frac{\lambda}{4!}\phi^4 +\cdots
\end{eqnarray}
with $m^2 \equiv -F(0)$.  

The nonlocal structure of (\ref{proto}) is encoded in the pseudo-differential operator $F(\Box)$.  In most cases of interest this can be represented as a
convergent series expansion
\begin{equation}
F(\Box) = \sum_{n=0}^{\infty} f_n \Box^n
\end{equation}
where the power $\Box^n$ is to be understood as $n$-fold composition of the d'Alembertian operator with itself.  By assumption the \emph{generatrix}
$F(z)$ is therefore an analytic function of the complex variable $z$.  The  Weierstrass factorization theorem ensures that we can write
\begin{equation}
\label{kinetic}
  F(z) = \Gamma(z) \prod_{a=1}^M (z-m_a^2)^{r_a}
\end{equation}
with $\Gamma(z)$ having neither zeros nor poles at finite $z$.  The propagator $G(p^2) \sim 1/F(-p^2)$ therefore has precisely $M$ poles at the locations
$p^2 = -m_a^2$, with the $a$-th pole being of integer order $r_a$.  

The field equation derived from (\ref{proto}) takes the form
\begin{equation}
\label{KG}
  F(\Box) \phi = \Vint'(\phi)
\end{equation}
The initial value problem associated with (\ref{KG}) has been studied in \cite{niky1,niky2} using the formal generatrix calculus, and also in
\cite{woodard,mulryne,gomis,woodard2,woodard3,gomis,calcagni} using alternative approaches.

\subsection{Multi-Particle Form of the Action}

Let us now re-write the prototype Lagrangian (\ref{proto}) in a form which is equivalent but which makes certain key features 
more transparent.  (See also \cite{nl_rev,pais}.)

We assume a kinetic function of the form (\ref{kinetic}) and perform the field redefinition\footnote{This is an allowed manipulation because
$\Gamma(z)$ contains neither zeros nor poles.} 
\begin{equation}
  \varphi = \Gamma(\Box)^{1/2} \phi
\end{equation}
so that the Lagrangian (\ref{proto}) becomes 
\begin{equation}
  \mathcal{L} = \frac{1}{2} \varphi \prod_{a=1}^{M}(\Box-m_a^2)^{r_a} \varphi - \Vint\left[\Gamma^{-1/2}(\Box)\varphi\right]
\end{equation}
Next, we introduce $M$ independent ``constituent'' fields defined as 
\begin{equation}
\label{constituent}
  \varphi^a = \prod_{b\not= a} (\Box-m_b^2)^{r_b} \varphi
\end{equation}
The ``collective'' field variable $\phi$ of the original theory is a superposition of these constituents:
\begin{equation}
\label{collective}
  \phi = \Gamma^{-1/2}(\Box) \sum_{a=1}^M \eta_a \varphi^a
\end{equation}
where the coefficients $\eta_a$ define a partial fraction decomposition of the quantity $H(z) \equiv F(z)/\Gamma(z)$ as
\begin{equation}
\label{eta}
  \sum_{a=1}^M \frac{\eta_a}{(z-m_a^2)^{r_a}} = \left[ \prod_{a=1}^M (z-m_a^2)^{r_a} \right]^{-1}
\end{equation}

Now, the Lagrangian (\ref{proto}) can be written in the decoupled form
\begin{equation}
\mathcal{L} = \sum_{a=1}^M \frac{1}{2} \eta_a \varphi^a (\Box-m_a^2)^{r_a} \varphi^a  
- \Vint\left[ \Gamma^{-1/2}(\Box) \sum_{b=1}^M \eta_b  \varphi^b\right]
\label{multi}  
\end{equation}
Equation (\ref{multi}) is the main result of this sub-section, it shows how the Lagrangian (\ref{proto}) for the ``collective'' field $\phi$
can be decomposed into a theory of $M$ ``constituent'' fields $\varphi^a$ (one for each pole of the propagator).   Notice that we have not made any particular assumptions 
about the geometry of the background space-time, nor have we employed perturbation theory.  In Appendix B we show how 
to use the multi-particle form of the action (\ref{multi}) to construct the perturbative Hamiltonian and look for constituent ghost 
excitations.

Using (\ref{constituent}), (\ref{collective}) and (\ref{eta}) is is possible to re-write the dynamical equation (\ref{KG}) in the form
\begin{equation}
\label{multi_KG}
  (\Box - m_a^2)^{r_a} \varphi_a = \Gamma^{-1/2}(\Box) \Vint'\left[\Gamma^{-1/2}(\Box) \sum_b \eta_b\varphi^b\right]
\end{equation}
(Notice that here $\Vint'$ is the derivative of $\Vint$ with respect to the collective field $\phi$, and \emph{not} with respect to the constituent
$\varphi^a$.)

\subsection{Dissipation and Delays}
\label{subsec:dissipation}

Our examples in section \ref{sec:string} both inherit their delayed structure from some source of dissipation in the dynamics.  We can model the
effects of dissipation on our prototype nonlocal QFT (\ref{proto}) is a very simple way, by adding a friction term to the d'Alembertian.  Thus, we take
\begin{equation}
\label{friction}
  \Box = -\partial_t^2 - \beta\partial_t
\end{equation}
when acting on a homogeneous function.  There are many ways to motivate the appearance such a friction term.  Obviously, we could obtain 
(\ref{friction}) by working in an expanding FRW universe.  In this case we would have $\beta = 3 H$ where $H =\dot{a}/a$ is the Hubble scale.
More generally, one might imagine motivating the appearance of such a term by supposing that the motion of $\phi$ leads to particle production,
in which case $\beta$ would be related to the imaginary part of the effective action.  (See, for example, \cite{KLS97}.)

To see how the simple-minded modelling of dissipation encapsulated in equation (\ref{friction}) can lead to delay-type nonlocality, let us assume
a kinetic function of the form (\ref{kinetic}) with
\begin{equation}
\label{gamma}
  \Gamma(\Box) = e^{-\alpha \Box}
\end{equation}
This type of nonlocality is ubiquitous in string theory; factors of the form $e^{-\alpha\Box}$ typically arise at the vertices
of Feynman diagrams.  Using (\ref{friction})
and (\ref{gamma}) we can re-write the constituent equation of motion (\ref{multi_KG}) in the form
\begin{equation}
\label{multi_almost}
  (-\partial_t^2 - \beta\partial_t - m_a^2)^{r_a} \varphi^a(t) 
  = e^{-\frac{\alpha}{2} \partial_t^2} \Vint'\left[ e^{-\frac{\alpha}{2}\partial_t^2} \sum_b \eta_b \varphi^b(t-\alpha\beta) \right]
\end{equation}
where we assume that $\beta$ is a constant.  We now introduce yet another set of basis fields\footnote{Once again, this manipulation is allowed 
because $e^{-\alpha s^2 / 2}$ has neither zeroes nor poles.}
\begin{equation}
  \psi^a(t) \equiv e^{-\frac{\alpha}{2}\partial_t^2} \varphi^a(t)
\end{equation}
Finally, the dynamical equation (\ref{multi_almost}) for the constituents takes the form 
\begin{equation}
\label{multi_delay}
  (-\partial_t^2 - \beta\partial_t - m_a^2)^{r_a} \psi^a(t) = e^{-\alpha\partial_t^2}\Vint'\left[ \sum_b \eta_b \psi^b(t-T) \right]
\end{equation}
where the delay is
\begin{equation}
  T = \alpha\beta
\end{equation}
This we suppose is positive.  Equation (\ref{multi_delay}) is the main result of this sub-section.

Equation (\ref{multi_delay}) is almost exactly of the form (\ref{dde}).  The only complication is the nonlocal operator $e^{-\alpha\partial_t^2}$ 
on the right hand side.  We can seek solutions by the method of steps.  (It might be that this approach does not access the full solution space.)  
Each constituent $\psi^a(t)$ is split into solution segments 
$\psi^a(t)=\psi^a_i(t)$ on the interval $t\in\left[(i-1)T,iT\right]$ with integer $i$.  (Here the upper index on the field denotes the constituent
label while the lower index labels the segment.)  The $i$-th solution segment can be obtained recursively as
\begin{equation}
\label{multi_step}
 (-\partial_t^2 - \beta\partial_t - m_a^2)^{r_a} \psi^a_i(t) = e^{-\alpha\partial_t^2}\Vint'\left[ \sum_b \eta_b \psi^b_{i-1}(t-T) \right]
\end{equation}
On the first segment we have an arbitrary initial function for each constituent field
\begin{equation}
\label{IFF3}
  \psi^a(t) = \psi^a_0(t) \hspace{5mm}\mathrm{on}\hspace{5mm}t\in \left[-T,0\right]
\end{equation}
Since equation (\ref{multi_step}) is $2r_a$-th in time derivatives, it follows that we can make $\psi^a(t)$ and its first $2r_a-1$ derivatives continuous at the knots $t=0,T,2T,\cdots$

It is worth pausing to admire the simplicity of equation (\ref{multi_step}).  We can construct solutions of a nonlinear differential equation of
infinite order simply by solving a set of $M$ inhomogeneous linear equations, each of which is finite order in derivatives.  The only complication is the appearance of the
nonlocal operator $e^{-\alpha\partial_t^2}$ in the source term.

In this section we have shown that the naive inclusion of dissipation (\ref{friction}) into the general prototype nonlocal theory (\ref{proto}) leads to
delay-type nonlocality, as manifested in the dynamical equation (\ref{multi_delay}).  Even though the nonlocality is not purely of the delay type, we are still
able to employ the method of steps.  In the next section 
we will show how this approach can be used to construct efficient and numerically stable time-dependent solutions.

Equation (\ref{IFF3}) provides a novel new way of thinking about the initial value problem for the nonlocal theory (\ref{proto}).  Each pole of the propagator contributes
a ``constituent'' field $\varphi^a$ with its own independent initial function.

Of course, the fact that a delayed formulation (\ref{multi_step}) with initial value problem (\ref{IFF3}) is possible does \emph{not} imply that
the theory (\ref{proto}) is stable or usefully predictive in any sense.  These key questions will depend sensitively on the underlying field theory
and the class of allowed initial functions.  It seems necessary to proceed on a case-by-case basis.  In the next section, we will consider a specific
example.

\section{A Case Study: Cosmological D-Brane Decay}
\label{sec:timelike}

\subsection{Delayed Formulation and the Method of Steps}

In the last section we showed how the simple-minded inclusion of dissipation leads to a delayed formulation of the dynamics in a very general
nonlocal QFT context.  Here we consider a specific example: D-brane decay in a background de Sitter space-time.  This toy model may be relevant
for nonlocal models of the early universe \cite{padic_inf,ng1,ng2} and also quintessence \cite{phantom}.  We employ the same level zero truncation
of SFT as was studied in subsection \ref{subsec:sft}, however, this time we take a constant dilaton profile and minimally couple $\phi$ to Einstein
gravity:
\begin{equation}
\label{SFT}
  \mathcal{L} = \frac{1}{g_s^2}\left[ \frac{1}{2}\phi C^{-2 \Box}\left(\Box + 1 \right) \phi - \frac{C^3}{3} \phi^3  \right]
\end{equation}
(see \cite{sft} for more details).  This example is consistent with our prototype model (\ref{proto}) and also the more restrictive class of theories 
considered in subsection \ref{subsec:dissipation}.   The equation of motion
\begin{equation}
\label{SFT_KG}
  C^{-2 \Box } \left( \Box + 1 \right) \phi = C^3 \phi^2
\end{equation}
admits infinitely many initial conditions and the Hamiltonian associated with (\ref{SFT}) is unbounded below \cite{niky1} (see Appendix C
for a review).  Our goal is to derive a delayed formulation of this equation and argue that the theory can be constrained to a stable subset of its
solution space at the expense of certain mild physical restrictions on the class of allowed initial functions.

In a background de Sitter space-time the equation of motion (\ref{SFT_KG}) can be written as
\begin{equation}
\label{delayed_SFT}
   \left(\partial_t^2 + 3 H \partial_t-1\right)\phi(t) = -C^{3-2\partial_t^2}\phi^2( t - T ), \hspace{5mm} T \equiv 6 H \ln C
\end{equation}
Following subsection \ref{subsec:dissipation} we can employ the method of steps.  The $i$-th solution segment $\phi_i(t)$ is constructed recursively as
\begin{equation}
\label{step_SFT}
  \left(\partial_t^2 + 3 H \partial_t - 1 \right)\phi_{i}(t) = - C^{3-2\partial_t^2} \phi_{i-1}^2( t - T )
\end{equation}
On the first segment ($i=0$) we have the initial function: $\phi(t)=\phi_0(t)$ on $t\in\left[-T,0\right]$.  Since the differential operator on the left hand side of
(\ref{step_SFT}) is second order, we can demand that the field and it's first derivative are both continuous at the points $t=iT$ where the solution segments
are joined.

\subsection{Allowed Initial Functions}

Which restrictions on $\phi_0(t)$ are necessary to constraint the theory (\ref{delayed_SFT}) to a stable subset of its solution space?  Following the 
analysis of subsection \ref{subsec:sft}, we might hope that the simple physical requirement $0 < \phi_0(t) < C^{-3}$ will be sufficient.  However, this
is not the case, we will also need to bound the derivatives of $\phi_0(t)$.

We can see which class of $\phi_0(t)$ will lead to late-time instability in a very simple way, by inspection of the delayed formulation (\ref{step_SFT}).
For a given initial function $\phi_0(t)$ we can expand the quantity $-C^3 \phi^2_0(t-T)$ on the interval $-T \leq t \leq 0$ as
\begin{equation}
\label{fourier_thingy}
  -C^3\phi_0^2(t-T) = \sum_n \alpha_n e^{2\pi i  n t / T}
\end{equation}
(For the sake of this argument the reader may interpret the summation over $n$ as either discrete or continuous.)
Now consider the $i=1$ solution segment, $\phi_1(t)$.  This obeys an inhomogeneous equation
\begin{equation}
  \left(\partial_t^2 + 3 H \partial_t - 1 \right)\phi_{1}(t) = \sum_n C^{8\pi^2 n^2 / T^2} \alpha_n e^{2\pi i n t / T}
\end{equation}
Then the dominant contribution to the source term obviously comes from the mode $n$ which maximizes the quantity $C^{8 \pi^2 n^2 / T^2} \alpha_n$.  Unless the
coefficients in (\ref{fourier_thingy}) tend to zero sufficiently fast, then it may happen that the dominant contribution to the source comes 
from some high frequency mode with $n \gg 1$.  If that happens then the particular solution of (\ref{step_SFT}) will go like $\phi_1(t) \sim e^{i\omega t}$ 
with $\omega \gg H$.  Carrying through the same procedure, the next segment oscillates even faster, as $\phi_2(t) \sim e^{2 i \omega t}$.  And so on.  At 
each successive step the energy cascades into higher-and-higher frequencies so that the solution will oscillate ever more rapidly.  This is precisely the 
usual Ostrogradski instability.  It is clear that in order to evade this pathological behaviour we must ensure that the Fourier coefficients $\alpha_n$ tend to 
zero sufficiently quickly at high frequency.  

Establishing a necessary condition on $\phi_0(t)$ for late-time nonperturbative stability is very difficult.  Therefore we will instead seek a sufficient 
condition.  An obvious candidate is the constraint $0 < \phi_0(t) < C^{-3}$ and also that $\partial_t^{(n)}\phi_0(t) / \phi_0(t) \lsim H^n$.  In the next section 
we will test this conjecture using fully nonlinear numerical simulations.  Although this constraint is slightly stronger than those which were considered
in section \ref{sec:string}, it is still far from onerous.  Indeed, we might imagine that this type of consistency condition could be established
dynamically within the context of a more realistic model of dissipation.

\subsection{Numerical Methods and Nonlinear Solutions}

We now implement numerically the method of steps (\ref{step_SFT}).  Obviously, to proceed we need to a faithful
numerical implementation of the infinite order operator $C^{-2\partial_t^2}$.  We could proceed by decomposing $\phi_{i-1}^2(t-T)$ into a Fourier 
series on the segment $t \in \left[(i-1)T,iT\right]$.  Then each member  of the sum is an eigenfunction of $\partial_t^2$ so that $C^{-2\partial_t^2}$ acts
in a simple way, term-by-term.  However, this idea turns out to be a very bad approach.  The problem is that the Gibbs phenomenon introduces errors at 
the points $t=iT$ where the solution segments are joined.  These errors propagate and become very large after just a few steps.  A much 
better approach is to represent the quantity $\phi_{i-1}^2(t-T)$ using an expansion in Chebyshev polynomials.  When this expansion is truncated to order 
$2 N$ (with $N$ integer) then the pseudo-differential operator $C^{-2\partial_t^2}$ is equivalent to the partial sum 
$\sum_{n=0}^{N} \frac{(-2\ln C)^n}{n!}\partial_t^{(2n)}$ and is straightforward to implement numerically.  

When the dissipation is sufficient to allow the field to settle down to the minimum at late times, then our iterative solution of (\ref{delayed_SFT}) is very efficient and
numerically stable, a significant improvement over higher order perturbation theory or the diffusion equation formulation (of course, our method is also
considerably less general than those alternative approaches).  These results are robust when the number of steps or the order of the Chebyshev 
polynomials is increased.  On the other hand, when the dissipation is small then the method breaks down within just a few steps.  However, since we 
regard wildly unstable solutions as unphysical, this is not a serious drawback.  

\begin{figure}[htbp]
\centerline{{\epsfxsize=0.5\textwidth\epsfbox{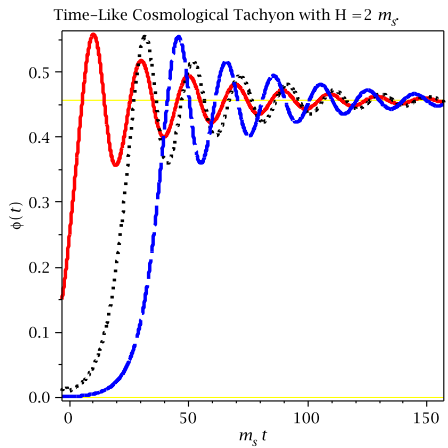}} }
\caption{The time-evolution of the time-like SFT tachyon in the presence of cosmological (de Sitter) friction for several representative choice of initial 
function, $\phi_0(t)$.  We set $H = 2 m_s$ for illustration.  Provided the constraint $0 < \phi_0(t) < C^{-3}$ is satisfied, and also that the derivatives of 
$\phi_0(t)$ are not too large, then the field will eventually settle down to the true minimum at late times.  The horizontal yellow lines illustrate the critical 
points $\phi=0$ and $\phi=C^{-3}$.   The explicit choices of $\phi_0(x^+)$ are provided in Appendix A.
}
\label{fig:timelike}
\end{figure}

We have investigated numerically the sufficient conditions on $\phi_0(t)$ for late-time stability.  The results confirm our previous intuition.  We will have
$\phi \rightarrow C^{-3}$ as $t\rightarrow \infty$ as long as $0 <\phi_0(t) < C^{-3}$ on $t \in \left[-T,0\right]$ and the derivatives of $\phi_0(t)$ are not too
large as compare to the coefficient of dissipation.\footnote{Quantitatively we can demand that the Laplace transform of $\phi_0(t)$ is analytic inside
a circular contour of radius $H$.  We stress that this is a sufficient, but not necessary, constraint.}  These dynamics are illustrated in 
Fig.~\ref{fig:timelike}.

Our results suggest that nearly \emph{any} source of dissipation will allow us to salvage sensible dynamics from the theory (\ref{SFT}), even without making 
restrictive assumptions about the symmetries of the solution.  We find this prospect encouraging since the rolling tachyon is supposed to be decaying 
into closed string excitations \cite{closed_emission} which were \emph{not} taken into account in the effective description (\ref{SFT}).  Thus, our results may provide a plausible 
mechanism by which the consistent dynamical inclusion of closed strings could allow SFT to evade the Ostrogradski instability.  At the more practical 
level, this general prescription helps to establish the viability of nonlocal cosmological models.

\subsection{Wild Oscillations and Allowed Initial Functions}
\label{subsec:wild}

It has been argued previously that the dynamics of tachyon condensation are stable for the special case where $\phi(t) \rightarrow 0$ as $t \rightarrow -\infty$.  For example, this 
was discussed in \cite{delays} for the light-like dilaton models discussed in section \ref{sec:string} (although in that paper the analysis went beyond our simple level-0 truncation).  See
also \cite{gomis}.  It is not clear if demanding that the field sit at an unstable maximum can be consistent with quantum mechanics (or, for that matter, macroscopic predictivity).  It
should therefore be emphasized that our class of ``allowed'' initial functions is much broader than this very special subset of solution space.  We only require certain qualitative restrictions
on the dynamics of $\phi(t)$ over a microscopic interval $\Delta t \sim m_s^{-1}$.  Moreover, our solutions depend on infinitely many free parameters and, 
in this sense, encode information about
the full nonlocal structure of the equations.  Nevertheless, it is worth comparing our analysis with the special case where $\phi(t) \rightarrow 0$ as $t \rightarrow -\infty$.  Because $\phi=0$ is a surface
of degeneracy, there is only a single growing mode which can be accessed by naive perturbation theory about the unstable maximum: $\phi_0(t) = \epsilon C^{-3} e^{r_+ t}$, with $\epsilon \ll 1$ and 
$r_+ = -\frac{3 H}{2} + \frac{1}{2}\sqrt{9 H^2 + 4}$.  Using the method of steps, we have confirmed numerically that this special class of initial function always leads to sensible late-time evolution, 
at least when $H \gsim 1$.  Notice that this special function is \emph{not} the unique perturbative solution of equation (\ref{delayed_SFT}) which satisfies  $\phi(t) \rightarrow 0$ as $t \rightarrow -\infty$.
Using the technique described in the appendix of \cite{BMNR} we could easily constuct a family of such solutions which depend on infinitely many free parameters.

In passing, it is worth noting that the wild oscillations of rolling tachyon solutions, which are a symptom of the Ostrogradsi instability, are not always viewed as pathological.
This comes down to a delicate issue of interpretation.  It is nontrivial to relate $\phi$ to physical observables in SFT.  Perhaps the ghostly constituent excitations are pure gauge.  In 
\cite{no_prob} it was argued that the unstable dynamics of $\phi$ can be interpreted as describing the closed string physics associated with Sen's tachyon matter \cite{tac_matter}.  
It is not clear to us how (or if) this claim addresses the issue of predictivity discussed in the introduction.  (See also \cite{gross,erler,taming} for related discussions.)  In this work our aim
is to address stability and predictivity in a general QFT context, without relying on intrinsically stringy physics.  Hence we regard any wildly unstable solutions as unphysical.

\section{Conclusions}
\label{sec:conclusions}

In this work we proposed a new formulation of the initial value problem for nonlocal theories with infinitely many derivatives.  This new approach is
less general than the pseudo-differential operator technique advanced in \cite{niky1,niky2}, however, it has considerable conceptual appeal.  In particular,
the underlying nonlocality of the model is manifest; the Cauchy surface has been ``smeared out'' over a finite width.

We have found that this mathematical re-formulation leads to a novel new perspective on the key problems of stability and predictivity which plague
nonlocal models.  By imposing mild physical restrictions on $\phi_0(t)$, it is possible to constraint the theory to a stable subset of its solution space.  
Moreover, the constrained theory makes robust predictions for dynamics on macroscopic scales.  Although our claims are limited only to a small number
of specific examples, we expect that similar results might be possible more generally.  It is our view that these restrictions on the initial function should
be viewed as a part of the definition of the theory.

Philosophically, our prescription is very similar to approaches which have been advocated previously for higher derivative theories of finite order.
Both the Cutkowski et al.\ contour deformation \cite{LW,lee-wick,cutkowski} and also the Hertog and Hawking \cite{hawking} approach are, classically, 
equivalent to adopting stability as a future boundary condition.  Both these approaches lead to a violation of microscopic causality, however, it is thought 
that macroscopic causality survives \cite{hawking,causal}.  In the initial function formulation we have an elegant and physically well-motivated way to 
impose late-time stability for models with infinitely many derivatives.  Our approach is nonperturbative and retains information about the full nonlocal 
structure of the theory.  Notice that microscopic causality is \emph{manifestly} violated in this formulation of the initial value problem.  We view this, 
along with the breakdown of microscopic predictivity, as a necessary feature of working with a fundamentally nonlocal theory.  In fact, these features 
seem rather natural: we should not \emph{expect} to be able to unambiguously resolve dynamics on times comparable to the underlying scale of 
nonlocality.

It is not our goal to claim that nonlocal models are generically sensible or phenomenologically viable.  Rather, we believe that fundamental nonlocality
is ubiquitous in well-motivated microscopic models and, therefore, it behooves us to try to understand such constructions from a general perspective.
By studying the connection between dissipation and delays, we have arrived at a new formulation of the initial value problem and a novel new 
perspective on the key issues of stability and predictivity.  These general lessons may be applied in a variety of contexts.  For example, the rolling
open string tachyon is, physically, supposed to be decaying into closed string excitations \cite{closed_emission} which are usually \emph{not} taken
into account in SFT.  Plausibly, the dissipation induced by this type of particle production has a dynamical effect which is similar to the dilaton gradient
in section \ref{sec:string} or the cosmological friction in section \ref{sec:timelike}.  Thus, we have elucidated a mechanism by which the consistent
inclusion of closed strings might enable SFT to escape the Ostrogradski instability.  A strength of our approach is that we do not rely on any intrinsically
stringy features of the underlying model.  Thus, our results may readily be applied to more general nonlocal QFT models (in particular, to nonlocal 
cosmological models).

There are a number of directions for future work on this subject.  Obviously, our results for nonlinear stability have been rather specific and 
model-dependent.  It would be very interesting to understand the connection between dissipation and delays in more general models, for example
where the coefficient of dissipation is time-dependent (as would arise in a self-consistent cosmology).  

Our analysis has been entirely classical, which is an obvious limitation.  Instabilities might re-appear at the loop level, or it might be that the field can
tunnel quantum mechanically to some unstable configuration even when the unstable directions in phase space are not classically accessible from an
allowed initial state (perhaps similarly to \cite{wesley}).  At the current level of understanding, it seems that perturbation theory is best 
suited to study quantization while the method of steps is best suited to study nonperturbative dynamics in the classical regime.  However, it might be 
possible to incorporate quantization directly into the delayed formulation.  One could imagine imposing 
$\left[\phi_0(t,{\bf x}),\dot{\phi}_0(t,{\bf x})\right]\not= 0$ on the initial segment $-T \leq t \leq 0$ and then using the method of steps to construct the 
q-number field $\phi(t,{\bf x})$ on $t > 0$.  It would be interesting to understand how the imposition of physical constraints on $\phi_0$ can be performed in a 
quantum theory.

One might also imagine trying to compute statistical properties of the $t>0$ dynamics by performing a functional average of the solution over allowed 
$\phi_0(t)$.  We find this idea fascinating, however, it is not clear how to put a measure on the space of initial functions.  We have investigated this 
idea using the exact solution (\ref{padic_soln}) where the averaging can be performed analytically.  In this particular case it seems that  the ambiguity in 
choosing the averaging procedure is not less sever than the ambiguity in choosing a realization of the initial function.  

It is interesting to note that our physical restrictions on $\phi_0(t)$ are not so stringent that they forbid violations of the Null Energy Condition (NEC).
In \cite{BMNR} this was shown explicitly for the toy $p$-adic model studied in section \ref{sec:string}.  Violations of the NEC play a key role
in cyclic/bouncing cosmologies, such as the new ekpyrotic model \cite{ekpyrotic}, but are notoriously difficult to obtain in a consistent microscopic
framework.  It is possible that our formalism could be used to construct sensible, NEC-violating nonlocal field theories.  We leave this, and other 
interesting possibilities, to future investigations.

\section*{Acknowledgments}

I am grateful to a number people for helpful discussions, correspondence and collaboration over the last few years which helped to 
shape the ideas in this text, including T.~Biswas, J.~Cline, N.~Kamran, N.~Moeller, S.~Patil, I.~Sachs, R.~Woodard and especially D.~Mulryne 
and N.~Nunes.   

\renewcommand{\theequation}{A-\arabic{equation}}
\setcounter{equation}{0}
\section*{APPENDIX A: Initial Functions for Numerical Analysis}

\renewcommand{\thesubsection}{A.\arabic{subsection}}
\setcounter{subsection}{0}

In this appendix we summarize the explicit initial functions used to generate the numerical results Fig.~\ref{fig:sft} and Fig.~\ref{fig:timelike}.  In the left panel
of Fig.~\ref{fig:sft} the solid red, dotted black and dashed green curves correspond, respectively, to the functions:
\begin{eqnarray*}
  \phi_0(x^+) &=& 10^{-2}\, e^{x^+} \\
  \phi_0(x^+) &=& 0.25 \, e^{x^{+} + T} \\
  \phi_0(x^+) &=& 0.25 \, e^{-3(x^+ + T)}
\end{eqnarray*}
In the right panel of Fig.~\ref{fig:sft} the solid red, dotted black and dashed green curves correspond, respectively, to the functions:
\begin{eqnarray*}
  \phi_0(x^+) &=& 0.1 \, \left[ 1 - 10 e^{-25 (x^++T)^2}\right] \, e^{x^+} \\
  \phi_0(x^+) &=& -10^{-2} \, e^{x^+} \\
  \phi_0(x^+) &=& 10^{-2} \, \sin\left[\frac{2\pi x^+}{T}\right] \, e^{x^+}
\end{eqnarray*}
In Fig.~\ref{fig:timelike} the solid red, dotted black and dashed green curves correspond, respectively, to the functions:
\begin{eqnarray*}
  \phi_0(t) &=& 0.25 \, e^{+r_+ t} \\
  \phi_0(t) &=& 10^{-2} \, \left[ 1 + 0.85 t^2 \cos\left(t / 2\right) e^{t/2}\right] \\
  \phi_0(t) &=& 10^{-3} \, e^{r_+ t}
\end{eqnarray*}
where $r_+ = -\frac{3H}{2} + \frac{1}{2}\sqrt{9H^2 +4}$, as defined in subsection \ref{subsec:wild}.

\renewcommand{\theequation}{B-\arabic{equation}}
\setcounter{equation}{0}
\section*{APPENDIX B: Conditions for a Bounded Hamiltonian}

\renewcommand{\thesubsection}{B.\arabic{subsection}}
\setcounter{subsection}{0}

In general the interactions in (\ref{multi}) will be nonlocal and the dynamics are very complicated.  We can gain some insight into the dynamics of
the theory by adopting a perturbative approach.  We write the each constituent field as 
\begin{equation}
  \varphi^a = \delta\varphi^a
\end{equation}
and expand the Lagrangian (\ref{multi}) order-by-order in perturbation theory
\begin{equation}
  \mathcal{L} = \delta\mathcal{L}^{(2)} + \delta\mathcal{L}^{(3)} + \cdots
\end{equation}
where the quadratic part $\delta\mathcal{L}^{(2)}$ gives the linearized equations of motion and the higher order contributions encode the effects
of interactions.  The quadratic Lagrangian decomposes into constituents as
\begin{equation}
\label{quadratic}
  \delta\mathcal{L}^{(2)} = \sum_{a=1}^{M}\delta \mathcal{L}^{(2)}_a, \hspace{5mm} 
\delta\mathcal{L}^{(2)}_a=\frac{\eta_a}{2}\delta\varphi^a (\Box-m_a^2)^{r_a}\delta\varphi^a
\end{equation}
Obviously the quadratic part of the Hamiltonian must also admit a similar decomposition
\begin{equation}
\label{H}
  \delta\mathcal{H}^{(2)} = \sum_{a=1}^M \delta\mathcal{H}^{(2)}_a
\end{equation}
where each $\delta\mathcal{H}_a^{(2)}$ is constructed from $\delta\mathcal{L}_a^{(2)}$.

Under what conditions can the perturbative Hamiltonian (\ref{H}) Hamiltonian be bounded below?  If $r_a > 1$ for any $a$ then $\delta\mathcal{L}_a^{(2)}$
must depend non-degenerately on time derivatives higher than than first order.  By the Ostrogradski theorem \cite{woodard2}, it follows that 
$\delta\mathcal{H}_a^{(2)}$  must be unbounded below.  Hence, only theories with $r_a=1$ for all $a$ yield a bounded Hamiltonian.  In this case the 
coefficient $\eta_a$ can be computed easily from the residues of the propagator at the pole $m_a^2$:
\begin{equation}
  \eta_a = \oint_{C_a} \frac{dz}{2\pi i} \frac{1}{H(z)} = \frac{1}{H'(m_a^2)} \hspace{5mm}\mathrm{where}\hspace{5mm}
	H(z) \equiv \frac{F(z)}{\Gamma(z)} = \prod_{a=1}^M (z-m_a^2)
\end{equation}
(The contour $C_a$ only encloses the pole at $z=m_a^2$.)  If any $\eta_a$ are negative, then (\ref{multi}) must contain ghosts.  
If $M>1$ then, because $H(z)$ is analytic, the residues must flip sign and some $\mathcal{H}_a$ will be unbounded below.  
Therefore the only possibility to construct a perturbatively ghost-free theory is when the propagator $G(p^2)\sim 1/F(-p^2)$ contains 
at most a single pole.

We have implicitly assumed that $m_a^2$ is real valued for all $a$.  The case of complex mass-squared is slightly more subtle since
now the $\eta_a$ coefficients are complex, as are the fields $\delta\varphi^a$.  Hermiticity of the action requires that such complex poles arise in conjugate 
pairs, so that if $m^2$ is a zero of $F(z)$ than so must be $(m^2)^\star$.  Let us suppose that $m_1^2$ and $m_2^2$ are complex conjugate pairs.  Then, 
reality of the solution implies that $\delta\varphi^2=(\delta\varphi^1)^\star$ and the relevant part of the Lagrangian is
\begin{equation}
  \frac{\eta_1}{2}\delta\varphi^1 (\Box - m_1^2) \delta\varphi^1 + \mathrm{c.c.}
  = - \frac{1}{2}K_{AB} \partial_\mu \delta\psi^A \partial^\mu \delta\psi^B - \frac{1}{2}M_{AB} \delta\psi^A\delta\psi^B
\end{equation}
where $\mathrm{c.c.}$ denotes the complex conjugate of the preceding term and in the second equality we
have defined $\delta\varphi_1 = \frac{1}{\sqrt{2}}\left(\delta\psi^1 + i \delta\psi^2\right)$ so that the indices $A,B$ run over real and imaginary
parts.  The kinetic term can be diagonalized by a rotation $\delta\psi^A \rightarrow R^A_B \delta\psi^B$.  It is easy to verify that one eigenvalues
of $K_{AB}$ is always negative and the Hamiltonian is unbounded below.

To summarize the results of this appendix: we have shown that a necessary (but not sufficient) condition for (\ref{proto}) to yield a 
perturbatively stable Hamiltonian is that the propagator $G(p^2)\sim 1/F(-p^2)$ has at most a single pole.  In this case the theory describes only a 
single degree of freedom and the equation of motion (\ref{KG}) requires only two initial data.\footnote{Stable theories describing multiple degrees
of freedom can be constructed when the generatrix is not analytic, for example if $F(z)$ has isolated poles or branch cuts.}  Absent the imposition
of some constraints (for example on the initial function $\phi_0(t)$ or following \cite{LW,lee-wick,cutkowski,hawking}) ghost-like constituent excitations
will lead to catastrophic classical instabilities and vacuum decay in the quantum theory.  See \cite{jim} for stringent observational constraints on the existence
of physical ghost fields.

There is one scenario where the analysis of  this appendix can give very misleading results: the case where $\phi=0$ is a surface of degeneracy
for the theory (\ref{proto}).  In this case the propagator obtained in a naive perturbation theory about $\phi=0$ will have less poles than the total number
of physical degrees of freedom in the theory.  This situation actually occurs for the level truncation of SFT, see appendix B for a review.

\renewcommand{\theequation}{C-\arabic{equation}}
\setcounter{equation}{0}
\section*{APPENDIX C: Counting Initial Data for the SFT Tachyon}

\renewcommand{\thesubsection}{C.\arabic{subsection}}
\setcounter{subsection}{0}

It is sometime incorrectly argued that
(\ref{SFT_KG}) admits only two initial conditions.  This specious conclusion is reached by perturbing about the constant solution $\phi=0$, which
corresponds to the unstable maximum of the potential.  Writing $\phi = 0 + \delta\phi$ we have 
\begin{equation}
\label{pert_wrong}
  C^{-2 \Box} \left( \Box +1\right) \delta\phi = 0
\end{equation}
to linear order in $\delta \phi$.  The generatrix is $F(z) = C^{-2 z}(z + 1)$ which has only a single zero corresponding to a tachyonic
excitation with mass-squared $M^2=-1$ (in string units).  At higher order in perturbation theory the counting of initial data is the same.  
However, this conclusion is very misleading.  Notice that perturbing about the \emph{other} critical point gives a \emph{different} answer.  Writing 
$\phi = C^{-3}+\delta\phi$ and linearizing in 
$\delta\phi$ we obtain
\begin{equation}
\label{another_pert}
  \left[ C^{-2\Box} \left(\Box + 1\right) - 2 \right] \delta \phi = 0
\end{equation}
The  generatrix is now $F(z)=C^{-2 z}(z + 1) - 2$ which has infinitely many zeroes $F(M_n^2) = 0$ given by
\begin{equation}
\label{Mn}
  M_n^2 = -\left[1 +  \frac{1}{2\ln C} W_n\left( -\frac{4 \ln C}{C^2} \right) \right]
\end{equation}
where $n=0, \pm1, \pm 2, \cdots$ and $W_n(x)$ denotes the branches of the Lambert-W function.  The spectrum (\ref{Mn}) corresponds to an 
infinite tower of constituent physical states with complex mass-squared.  Correspondingly, equation (\ref{another_pert}) admits infinitely many 
initial conditions.  In the Appendix of \cite{BMNR} we showed how to construct solutions of (\ref{SFT_KG}) near $\phi=0$ which depend on infinitely
many free coefficients.  Using the results of Appendix A, it follows that the Hamiltonian associated with (\ref{SFT}) must be unbounded below.
 
Perturbing about different critical points of the potential we found different numbers of initial conditions.  This is because $\phi=0$ is a surface of 
degeneracy for the equation of motion (\ref{SFT_KG}) \cite{woodard}.  This phenomenon has nothing to do with the nonlocal structure of the 
theory.\footnote{A simple example of  a local theory with the same property was provided in \cite{woodard}.  Consider the Lagrangian 
$\mathcal{L} = -\frac{M^3}{2}\psi \ddot{\psi}^2$.  The Euler-Lagrange equation is fourth order in time 
derivatives and hence generic solutions admit four initial data.  However, on the surface $\psi=0$ the character of the equation of motion changes: 
instead of determining $\partial_t^4\psi$ in terms of lower order derivatives it instead becomes a constraint on the lower derivatives.
Thus, solutions which start at the point $\psi=0$ contain only three free coefficients.}  

The fact that (\ref{SFT}) allows for negative kinetic energy cannot be doubted.  Time-dependent solutions in flat space have been shown to move 
\emph{higher} on the potential than their starting point, without difficulty \cite{zwiebach}.  Since energy is conserved in this rolling process, it follows that 
the kinetic energy \emph{must} be negative.  Moreover, generic solution of (\ref{SFT_KG}) undergo erratic oscillations with divergent frequency and 
amplitude, the hallmark behaviour of an Ostrogradski-sick theory.  The erratic  time dependence of rolling tachyon solutions is properly understood as 
the \emph{symptom} of a much deeper problem with SFT.

\end{document}